\documentclass[11pt,a4paper,twocolumn]{article}
\usepackage[utf8x]{inputenc}
\usepackage[T1]{fontenc}
\usepackage{mathptmx} % Use Times Font
\usepackage{amssymb} % for ticks (\checkmark)
\usepackage{amsmath}

\usepackage[draft]{changes}          % Then load changes package
\usepackage{algorithm}
\usepackage{algpseudocode}

\usepackage{soul}
\usepackage{xcolor}

\sethlcolor{yellow}

\newif\ifdraft
\draftfalse   % final version

\ifdraft
  \newcommand{\del}[1]{\sout{#1}}
  
\else
  \newcommand{\del}[1]{}
  
\fi

\renewcommand{\highlight}[1]{#1}       % highlighting OFF

\usepackage[normalem]{ulem} % provides \sout

\usepackage[normalem]{ulem}

\usepackage{soul}
\usepackage{xcolor}
\sethlcolor{yellow}

\soulregister\cite7
\soulregister\ref7
\soulregister\eqref7
\soulregister\footnote7

\definechangesauthor[name={nick}, color=orange]{nick}

\usepackage{caption}
\usepackage{tabularx}

\usepackage{graphicx} % Required for including pictures
\usepackage[pdftex,linkcolor=black,pdfborder={0 0 0}]{hyperref} % Format links for pdf
\usepackage{calc} % To reset the counter in the document after title page
\usepackage{enumitem} % Includes lists
\usepackage[mathscr]{euscript}
\usepackage{multirow}

\usepackage[a4paper, lmargin=0.11\paperwidth, rmargin=0.11\paperwidth, tmargin=0.1\paperheight, bmargin=0.1\paperheight]{geometry} %margins

\usepackage[all]{nowidow} % Tries to remove widows
\usepackage[protrusion=true,expansion=true]{microtype} % Improves typography, load after fontpackage is selected

\usepackage{mwe}%for title page
\hypersetup{ 
pdftitle = {Overcoming Labelled Data Scarcity for Defect Classification in Scanning Tunneling Microscopy},
pdfauthor = {Nick Kolev}
}

\title{Overcoming Labelled Data Scarcity for Defect Classification in Scanning Tunnelling Microscopy}

\begin{document} 
\begin{onecolumn}
\begin{titlepage}
	\centering
	{\LARGE \textsc{University College London}\par}
	\vspace{1cm}
	{\huge\bfseries \highlight{Overcoming Labelled Data Scarcity for Defect Classification in Scanning Tunneling Microscopy}\par}
	\vspace{2cm}
        {\Large \textbf{Nikola L. Kolev}$^{1,2,*}$, \textbf{Max Trouton}$^{1,4}$, \textbf{Filippo Federici Canova}$^{5,6}$, \\ \textbf{Geoff Thornton}$^{1,4}$, \textbf{David Z. Gao}$^{3,5,7}$, \textbf{Neil J. Curson}$^{1,2}$, \\ \textbf{Taylor J. Z. Stock}$^{1,2,*}$\par}
	\vspace{0.5cm}
	{\normalsize $^{1}$London Centre for Nanotechnology, \\
                           University College London,
                           17-19 Gordon Street, London, WC1H 0AH, UK\\
                  $^{2}$Department of Electronic and Electrical Engineering, \\
                  University College London, \\
                  London, WC1E 7JE, UK \\
	           $^{3}$Department of Physics and Astronomy, University College London, 
               London, WC1E 7JE, UK \\
                  $^{4}$Department of Chemistry, University College London, \\
                  London, WC1E 7JE, UK\\
                  $^{5}$Aalto Science Institute, School of Science, Aalto University, 02150 Espoo, Finland \\        
                  $^{6}$Nanolayers Research Computing LTD, \\
                  London, UK \\
                  $^{7}$Department of Physics, \\
                  NTNU Norwegian University of Science and Technology, \\
                  Trondheim, Norway \\
                  $^{*}$ Authors to whom any correspondence should be addressed using the emails: nick.kolev@hotmail.com, t.stock@ucl.ac.uk \\
                  \par}
	\vspace{2cm}

         % abstract is 191 words (16/04/2025)
        \begin{abstract}
           Scanning tunnelling microscopy (STM) is a powerful technique for imaging surfaces with atomic resolution, providing insight into physical and chemical processes at the level of single atoms and molecules. A regular task of STM image analysis is the identification and labelling of features of interest against a uniform background. Performing this manually is a labour-intensive task, requiring significant human effort. To reduce this burden, we propose an automated approach to the segmentation of STM images that uses both few-shot learning and unsupervised learning. Our technique offers greater flexibility compared to previous supervised methods; it removes the requirement for large manually annotated datasets and is thus easier to adapt to an unseen surface\del{while still maintaining a high accuracy}. We demonstrate the effectiveness of our approach by using it to recognise atomic features on three distinct surfaces: Si(001), Ge(001), and TiO$_2$(110), including adsorbed AsH$_3$ molecules on the silicon and germanium surfaces. Our model exhibits \highlight{potential for generalisation}, and following initial training, can be adapted to unseen surfaces with as few as one additional labelled data point. This work is a significant step towards efficient and material-agnostic, automatic segmentation of STM images.
        \end{abstract}

	\vfill

% Bottom of the page
	{\large \par}
\end{titlepage}
\end{onecolumn}
\twocolumn

\newpage

% Document is 5779 words (16/04/2025) not including fiugre captions and abstract.

\section{Introduction}
A scanning probe microscopy (SPM) image is a dense n$\times$n matrix of values, recording the point wise interaction between an atomically sharp probing tip and the atoms on a surface. In the case of scanning tunnelling microscopy (STM) \cite{Si(111)_atom_res}, the measured tunnelling electron current between tip and surface relates directly to the tip-sample separation and the local density of electronic states of the surface. An STM image therefore contains both topographic and electronic information. Given this complexity, the quantitative analysis of STM results requires correlation between the images and physical models of the measured system. In practice, this is achieved by comparing the results of theoretical studies with significant quantities of labelled STM data. Computer vision, the ability for computers to derive information from images, can be used to automate much of this procedure, and machine learning (ML) based image recognition finds a natural application in the analysis and segmentation of STM images. 

In existing work, Rashidi et al., and Ziatdinov et al. have both successfully trained convolutional neural networks (CNN) to detect a variety of known defects in STM images of the hydrogen terminated silicon surface \cite{ziatdinov2020} \cite{Rashidi_2020}. Additionally, the exploration of machine learning (ML) for real-time image improvement by automated STM tip monitoring, modification, and repair, has shown promising initial results and bodes well for the broader application of ML for task automation in the STM community \cite{rashidi_tip} \cite{OGordon_tip}  \cite{deepspm} \cite{gordon2020embedding}. Despite these successes, a notable limitation of existing applications of ML to STM is its reliance on foreknowledge of the measured surfaces. STM is regularly applied to new materials and novel experiments inherently give rise to novel data. However, accurate analysis using traditional supervised machine learning techniques, such as those used in \cite{Rashidi_2020} \cite{rashidi_tip} \cite{OGordon_tip}, requires significant amounts of labelled training data for each newly studied surface system. For example, the classification problems in references \cite{rashidi_tip} and \cite{OGordon_tip} use around 3000 labelled STM images, but training data sets of this size may may not be practical or even possible in other STM applications. This limitation can reduce or even remove the advantages offered by using ML for data labelling. 

One possible solution to this problem is the artificial synthesis of labelled data. This can, in principle, be accomplished with density functional theory (DFT) \cite{STM_DFT_simulated_scans}, but in practice this approach often requires the introduction of additional neural networks to address dissimilarities between theoretical and experimental data \cite{STM_cycleGan} \cite{STM_noiseremoval} \cite{atomAI}, not to mention the need for an expert in DFT calculations and many hours of computational resources. \highlight{Less expensive alternatives, such as tight-binding models, are available but require existing parameters for the target material and prior validation of their accuracy \cite{blanco2004first, korventausta2009stm}.} A more straightforward solution, to the onerous demand for extensive labelled data, is an ML image recognition technique for STM that relies on little or no data from unseen material systems. \highlight{Several recent studies have explored such approaches \cite{zhu2022deep, yuan2023applying}. For example, Wei et al. trained a ResNet-101 classifier on a dataset of approximately 240 molecules (around 40 images per class), relying heavily on data augmentation to expand their dataset \cite{wei2025two}. Although this represents only an order-of-magnitude reduction in dataset size compared to earlier work, it is nevertheless a meaningful step toward data-efficient STM analysis. Hellerstedt et al. proposed a workflow that avoids ML entirely by using Zernike polynomial features for molecular identification \cite{hellerstedt2022counting}. Their approach requires as few as one labelled image per class, although they do not report error analyses or performance metrics that would allow a rigorous comparison. In parallel, there have been efforts in the broader microscopy community to adapt the Segment Anything Model (SAM) of Kirillov et al.-a foundational vision model intended to generalise “out of the box” across diverse imaging domains \cite{kirillov2023segment}. However, work by Archit et al. indicates that SAM requires domain-specific finetuning to achieve competitive performance on microscopy data, and its vision-transformer architecture entails substantial computational cost \cite{SAMmicroscopy}.} In the approach presented here, we minimise the number of labelled examples required from each new experiment by using a combination of unsupervised learning and few shot learning (FSL) \cite{FSL_review}. \highlight{Our goal is to evaluate a range of FSL algorithms and identify those best suited to STM data, rather than search for the best architectures.}

FSL is a widely used image recognition technique that has also been adapted for other modalities where human discernment typically excels, such as audio, language, and radar \cite{heggan} \cite{proto_clip} \cite{radar_fsl}. Rather than learning to classify an object into several predetermined classes, FSL networks learn what it means for two objects to be the same. Training of an FSL model focuses on differentiating between two or more inputs, rather than classifying them, and this lifts the requirement for large training datasets. While initial training can still require large amounts of data (a few hundred labelled data points in our work), once completed, new problems can be successfully tackled with as little as one labelled data point per class \cite{FSL_review}. 

Despite its flexibility, to our knowledge, FSL has not yet been utilised in STM or in the broader SPM community. Notably, it has gained attention within the astronomy community for classifying galaxy morphologies \cite{galaxy_FSL1}, an arguably similar task to the one posed by STM data. FSL has also seen recent application in electron microscopy where it has been used for coarse segmentation of scanning transmission electron microscopy (STEM) images of epitaxial heterostructures \cite{akers_prototyp_SEM}, and space group classification of electron backscatter diffraction (EBSD) patterns \cite{transferlearing_EBSD}. In addition, although not strictly FSL, the MicroNet microscopy dataset, has been used to demonstrate accurate electron microscopy image segmentation using decoders trained on only a few fully segmented images \cite{MicroNet}. These successes in related fields suggest that FSL approaches may also be beneficial when applied to STM data.

In our work, we apply FSL to the STM study of atomic-scale defects on semiconductor surfaces. This is motivated by the technological importance of dopant precursor chemistry on the Si(001) and Ge(001) surfaces, which has applications in quantum computing \cite{PH3_on_Si} \cite{Si(001)_diethyl_ether}. When paired with hydrogen lithography (an atomically precise STM-based fabrication technique) \cite{lyding}, suitable dopant precursors allow for the creation of atomic-scale quantum electronic devices, including dopant-spin based qubits \cite{NIST_SETs} \cite{2_qubit_gate_P}. To date, single atom devices have been fabricated in this way using phosphine and phosphorus in silicon, but new precursors for dopants such as arsenic and boron have been established \cite{AsH3_on_Si} \cite{Boron_HLith1} \cite{BoronHlith2}. The STM experiments that are used to assess candidate precursors, examine the chemistry of new molecules on known surfaces. Having an ML tool that can easily adapt to STM data of new precursors and dopants, continually building upon the existing knowledge from past experiments, will help to speed the process of discovering suitable new precursors.

Compared to the examples of FSL in electron microscopies cited above, we not only develop an effective FSL approach for our test case STM data, but we also assess the suitability of different FSL algorithms for STM data analysis. Here, we investigate four different FSL algorithms: prototypical \cite{prototypical_net}, matching \cite{matching_net}, relation \cite{relation_net}, and simple shot \cite{simpleshot_net}, and find the prototypical network performs best, overall, with up to 99\% accuracy in our test cases. \del{Finally,} We compare pretrained networks with networks trained in-house on subject specific data and explore how the number of classes in a training set can improve the accuracies of our networks. \highlight{Finally, we also use kNN on the bare pixels as a baseline.} By evaluating several of the most popular FSL networks in this way, we aim to provide a guide for others in the SPM community and related fields who wish to apply FSL to their microscopy data. 

\section{Background}
\begin{figure*}[!htb]
    \centering
    \includegraphics[width=16cm]{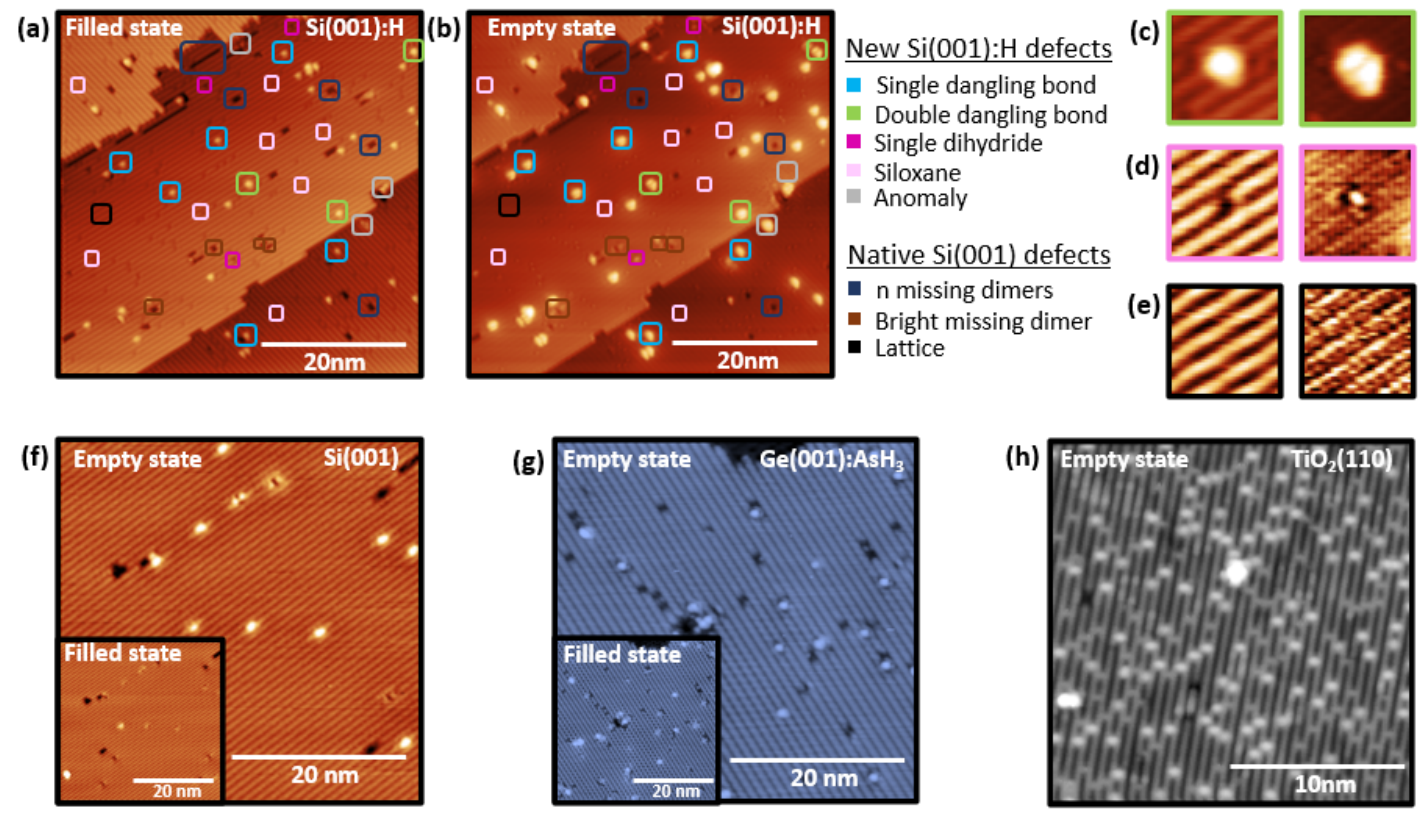}
    \caption{ (a) Filled and (b) empty state STM images of the same area of a Si(001):H surface. Imaging parameters: -2 V/+2 V, 50 pA. Different defect types are circled with different colours. Not all defects have been labelled to avoid overcrowding. (c)-(e) show cropped features from (a) and (b). (c) A double dangling bond (i.e. an exposed bare silicon dimer on the Si(001):H surface). (d) A siloxane feature (oxygen atom bridging a silicon dimer: Si-O-Si). (e) Detail of the hydrogen terminated silicon dimer rows. (f) Filled and empty state STM images of the Si(001) surface. Imaging parameters +2 V/-2 V, and 50 pA. (g) Filled and empty state images of the Ge(001):AsH$_3$ surface. Imaging parameters: -1.5 V/+1.0 V, 200 pA (h) Empty state STM image of the TiO2 (110) surface. Imaging parameters: +1.6 V, 100 pA. All images are plane-levelled and scan line aligned. 
}
    \label{explanation}
\end{figure*}

One routine application of STM is the exploration of surface chemical reactions. In these experiments, STM data reveals atomic scale features present on a surface before and after exposure to adsorbates. The larger the area one examines with the STM, the higher the statistical significance of the findings. Therefore, from an image recognition perspective, the task at hand is to identify and label all atomic-scale features in many similar images. As a specific example of this, Figures \ref{explanation}(a) and (b) present filled state (tunnelling from the sample) and empty state (tunnelling to the sample) STM images of a hydrogen terminated silicon surface: Si(001):H. 

When working with a known substrate and a new adsorbate, such as Si(001) and the terminating hydrogen in our example, the features will broadly divide into two classes: known native defects of the substrate, and new features attributed to the adsorbate. Traditionally, an STM user would sift through the data, labelling every instance of previously unseen defects and then attribute these to the new chemical species. This is illustrated in Figures \ref{explanation}(a) and \ref{explanation}(b), where coloured boxes are used to label native defects such as bright and dark silicon atomic vacancies, as well as features associated with the hydrogen termination, such as single and double dangling bonds and siloxane (adsorbed oxygen from background water). Examples of a double dangling bond and a siloxane feature are shown in detail in Figures \ref{explanation}(c), and \ref{explanation}(d), respectively \cite{Si(001)-H_features}.The identified features are distinguished from a uniform background, and in the case of Si(001):H, this background is composed of atomic dimer rows, shown in detail in Figure \ref{explanation}(e). Finally, in addition to identifiable feature types, there may also be anomalies present. These are defects that do not fit into any class, and which appear only once or twice in the data. Due to their scarcity, anomalies are usually dismissed during counting or included in uncertainties. An ML based approach to labelling data of this sort should be capable of handling both broad feature classes (known substrate defects and new adsorbate features) plus any anomalies.

To demonstrate our FSL based STM image recognition approach, we first train the networks using data from silicon surfaces, that is, Si(001):H shown in Figure \ref{explanation}(a) and (b), both with and without adsorbed AsH$_3$ molecules, plus data from the bare Si(001) shown in Figure \ref{explanation}(f). With the models trained, we then address two additional surfaces: arsine adsorbed germanium: Ge(001):AsH$_3$, and titania: TiO$_2$(110), examples of which are shown in Figures \ref{explanation}(g) and \ref{explanation}(h), respectively. While germanium’s structure and semiconductor properties are very similar to those of silicon, the well-studied titania surface \cite{TiO2_1} \cite{TiO2_2} \cite{TiO2_3} is different from the group IV semiconductor surfaces, in both appearance and structure, so that our three test cases represent a gradual progression from familiar to unfamiliar.

\section{Method}
\subsection{Image Recognition Workflow}

\begin{figure*}[!htb]
    \centering
    \includegraphics[width=15cm]{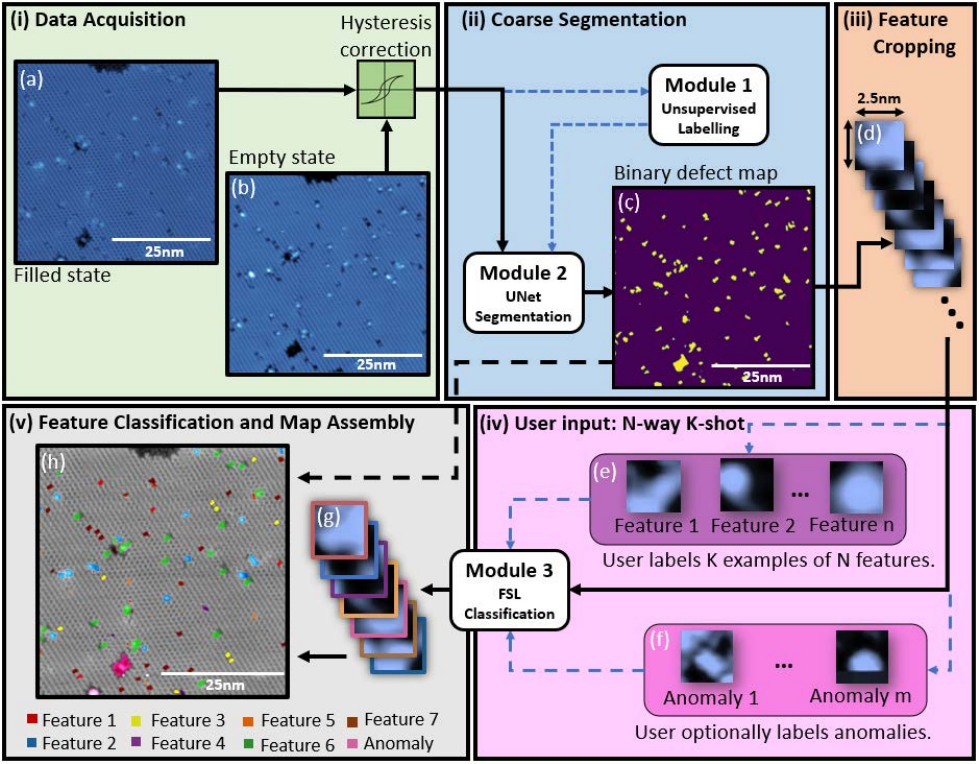}
    \caption{ STM Image Segmentation and Feature Classification Workflow: The workflow is illustrated using filled and empty state data, (a) and (b), from the Ge(001):AsH$_3$ system. It is divided into 5 steps (i through v) and utilizes three image processing modules (1 through 3). Step (i) STM data acquisition and preprocessing including hysteresis correction \cite{zhang2019method}. (ii) Course segmentation by a U-Net (Module 2) to produce a binary map (c). The U-Net is trained on data labelled using an unsupervised clustering technique (Module 1). (iii) Atomic defects identified in step ii are isolated in uniform cropped images (d) and resampled to match network dimensionality. (iv) The FSL support set (e) is generated by the user who defines N feature classes and labels K examples for each class using a subset of crops from step (iii). The user may also exclude anomalous features (f) from analysis at this step. (v) Using the support set defined in step (iv), the FSL network (Module 3) sorts all remaining unseen cropped features into the N classes. Finally, classified features of interest (g) are combined with the binary map (c) into the final output of an STM image feature map (h).}
    \label{workflow}
\end{figure*}

Figure \ref{workflow} presents an overview of our approach to FSL-based image recognition for STM data segmentation. Here we use the Ge(001):AsH$_3$ system of Figure \ref{explanation}(f) as an example. The workflow consists of five steps. \\

\textbf{(i) Data acquisition and preprocessing} \\
STM images are acquired in two channels (filled and empty states), \del{then distortion corrected, aligned, and} \highlight{plane levelled, scan line aligned, and then hysteresis is corrected (using an algorithm developed by Zhang et al. \cite{zhang2019method}). Images are} combined in image tensors that are fed forward through the image processing framework. \highlight{A full summary of all pre processing steps (including during training, and details such as the plane levelling algorithm used) are given in Appendix~\ref{app: preprocessing}.}\\
\textbf{(ii) U-Net binary segmentation} \\
The unseen data is coarsely segmented by a U-Net, each pixel is labelled either as defect or lattice. The U-Net module requires training data for each new substrate, and this is provided using an unsupervised learning module for data labelling, described below in section 3.3.\\
\textbf{(iii) Feature cropping} \\ 
The atomic defects identified in step (ii) are then isolated in uniform image crops that are resampled to match network dimensionality, before feeding forward to the FSL module. \\ 
\textbf{(iv)  User defined N-way K-shot problem} \\
 The user defines N feature classes, and labels K examples for each class using images from the set of crops collected in step (iii), thus generating the support set required by the FSL network. Within this step, the support set can be changed from one set of images to another, and anomalies may be excluded from analysis as the STM experiment progresses and the character of the data develops. \\ 
\textbf{(v) FSL classification} \\
Finally, with the support set established, the FSL network runs inference on the remaining unseen cropped features, sorting them into the N user defined classes. Once all features of interest have been classified, this information is combined with the course segmentation of step (ii), producing the final output, a segmented image with full classification of all atomic scale features of interest. \\

\subsection{Module 1: Unsupervised Labelling}

\begin{figure*}[!htb]
    \centering
    \includegraphics[width=15cm, height=7cm]{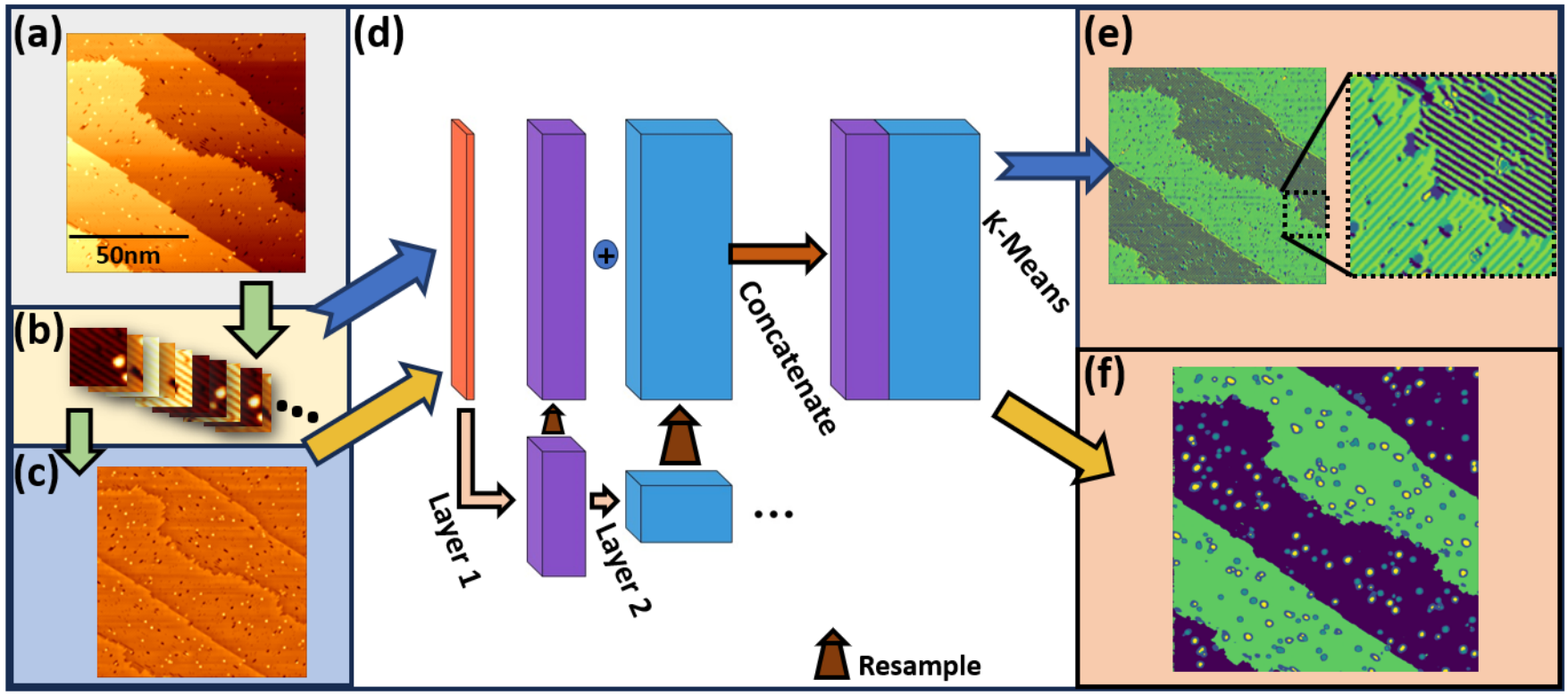}
    \caption{   
    Overview of the unsupervised pipeline for STM image segmentation. (a) Original filled-state STM image of the Si(001):H:AsH$_3$ surface (100 nm × 100 nm, 512 × 512 pixels; acquired at -2 V, 50 pA). (b) The image is partitioned into 64 × 64 pixel patches with 32-pixel overlap. Each patch undergoes normalisation (min-max followed by z-score). (c) Reconstructed image after applying a sinusoidal window function to the patches in (b). This serves as one potential input pathway. (d) Feature extraction. The network incorporates resampling layers (indicated by 'Resample' arrows) to process features at different scales. Inputs are normalized (min-max, z-score) as is needed by the FCNResnet101 network. (b),(d),(e) Pathway 1: High-resolution segmentation. Patches from (b) are processed through the network, yielding 768-dimensional feature vectors per pixel. Subsequent K-means clustering (K=7) produces the final segmentation map shown in (e), with a zoomed-in region highlighting detail. (c),(d),(f) Pathway 2: Lower-resolution segmentation. The pre-processed image (c) is fed through the network. K-means clustering (K=5) is applied to the extracted features to generate the segmentation map (f).}
    
    \label{auto_segment}
    
\end{figure*}
To produce accurate coarse segmentations in step (ii)  above we require multiple new labelled scans to train the U-Net for new substrates. \highlight{It would be possible to train a cross-substrate U-Net to improve generalisability across different materials. However, we consider this an avenue for future work. Within the scope of the present study and given the available data, we have focused on substrate-specific U-Nets, which allow us to demonstrate and validate the broader applicability of our approach.} Although our results demonstrate that as few as five scans can be sufficient, this still takes considerable time for a single person to annotate. To significantly reduce this burden, we utilize unsupervised learning to produce the labelled training data for the U-Net. Our approach here is similar to that developed by Archit et al. \cite{SAMmicroscopy}, using large pretrained models. The key difference is that we do not use the output of the pretrained models to fine-tune a new large model but instead train a smaller model capable of segmenting images on a CPU in seconds - in effect distilling the knowledge of the large, generalised model to our smaller, subject specific model.  \\

In our approach, we use the FCNResNet101 network \cite{FCN}\cite{Resnet101} pretrained on ImageNet to extract feature vectors for each pixel in an image, by taking the output of the first and second layers of the network (since the embeddings at these stages are less specialised to the ImageNet training data) and up sampling these to equal the resolution of the input image. We then use k-means clustering on the pixels' feature vectors to segment the image. \highlight{We hand tune the k-values for appearance such that they represent physical features such as atomic terraces or defects.} Here, we can vary the segmentation granularity by controlling resampling of the image before feature extraction. Feature extraction at lower resolution allows us to extract larger scale information such as crystal phase domain shape, size, and location, each of which is useful in STM studies in its own right \cite{PhaseSep1} \cite{PhaseSep2}. An example of this is presented in Figure \ref{auto_segment}(f), where the stepped Si(001) surface in Figure \ref{auto_segment}(a) is segmented into two domains (dimer rows rotated by 90 degrees separated by atomic steps), identified by the purple and green regions. Feature extraction at higher resolution, as shown in \ref{auto_segment}(e), leads to more detailed segmentation with dimer rows and defects both being identified, but not yet clustered into meaningful classes. From this more granular segmentation, we can extract a binary map of the location of all defects. An example is shown in the yellow overlay of Figure \ref{auto_segment}(f). This process requires a graphics processing unit (GPU), taking up to five minutes for an image upsampled to 4160$\times$4160 pixels on a Tesla T4 GPU. The slow computation and requirement for GPU access restricts the use of this method on the fly, during data acquisition, therefore the binary maps are used to train a U-Net for each new substrate, which can produce an output in seconds on a CPU. A faster, but less accurate, alternative is presented in Appendix \ref{auto_label_knn}.

\subsection{Module 2: U-Net Binary Segmentation}\label{UNet_sec}

Having produced labelled images for a particular substrate using module 1, we then train a U-Net \cite{UNet}, for use in module 2. This approach not only allows us to run our technique on a CPU but also allows for data augmentation to train a more robust network. We use a standard U-Net architecture with three down sampling layers, a bottleneck, and three up sampling layers. As with our FSL, we do not run a hyperparameter exploration. An example of the binary mapping, via the U-Net, is shown in Figure \ref{workflow}(c), which maps images \ref{workflow}(a) and (b).

In training for our three substrate test cases, we use five scans each of of 512$\times$512 pixels resolution and (100 nm)$^2$, (50 nm)$^2$, and (10 nm)$^2$ in size for Si, Ge, and TiO$_2$, respectively. We then apply data augmentation to deterministically add simulated experimental noise and STM tip artefacts to the scan data, and to increase the effective area of our scans by taking random crops of 64$\times$64 pixels and rotating randomly by 0, 90, 180, or 270 degrees. Using crops in this way also teaches the network translational and rotational symmetry, which can be tailored to the symmetries of the surfaces studied. Finally, when making predictions about an unseen image, a method similar to Pielawski et al. is used \cite{windowing_segmentation}. We slice the image into overlapping crops of equal size to the random crops used in training, i.e. 64$\times$64 pixels. The U-Net then segments each crop, and these are reassembled to give the final output. Overlapping the crops helps to reduce edge effects from the model \cite{edge_effects}. 

\subsection{Module 3: FSL Feature Classification}

\begin{figure*}[!htb]
    \centering
    \includegraphics[width=15cm]{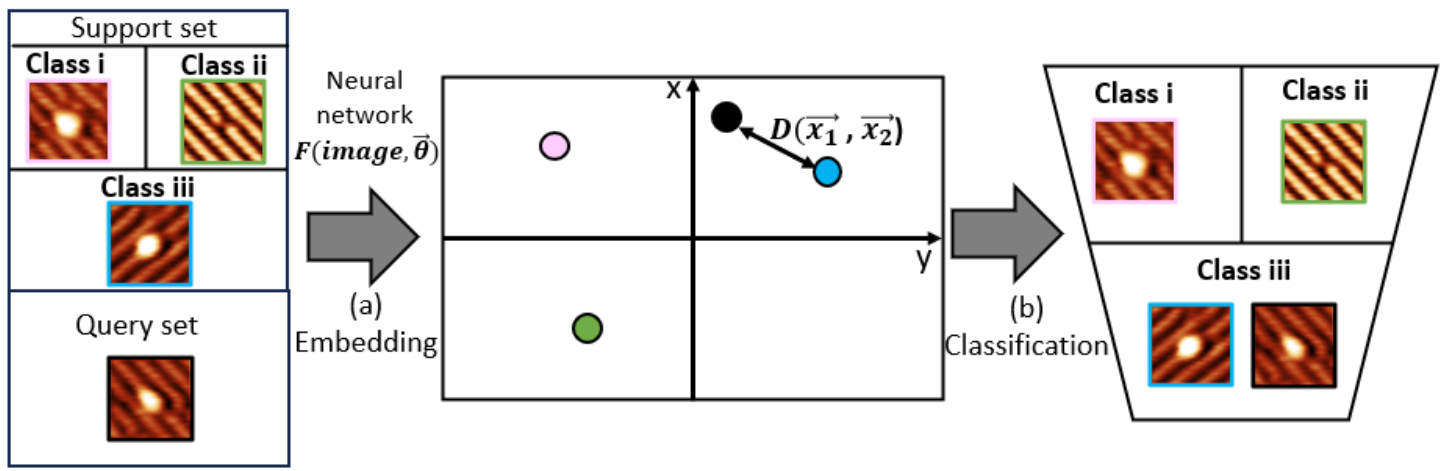}
    \caption{ Illustration of a metric based approach to few-shot learning. $F(image, \vec{\theta})$ and $D(\vec{x_i}, \vec{x_j})$ vary depending on the chosen approach. $\vec{\theta}$ are the weights of the neural network. The vector space is referred to as the latent space. Once images are (a) embedded in the vector space, they are (b) classified by determining the closest data point from the support set using $D(\vec{x_i}, \vec{x_j})$ as the metric.}
    \label{fsl_explain}
\end{figure*}

Few-shot learning is an ML technique to be used when labelled data is scarce. In this approach, the goal is to accurately classify unseen data using only a few examples of the data classes. FSL is generally described as an N-way K-shot classification problem. In each FSL problem, or episode, a support set, S, contains K labelled examples from N defined classes, and is used to infer the classification of an unseen query set, Q, which contains an equal distribution of samples to be classified into the N classes. An equal distribution is needed during training and testing to prevent over or under sampling and provide a fairer measure of accuracy but is not necessary during inference. At a higher level, as with supervised ML, there is still a training, test, and validation set: T$_{tr}$, T$_{te}$, and V, respectively. Crucially, since the goal is to train a network that can adapt well to new data, T$_{tr}$, T$_{te}$, and V cannot share any classes. In training, S and Q are generated by randomly sampling from T$_{tr}$, in testing from T$_{te}$, and in validating from V. In this paper, we focus on the metric-based approaches to FSL.

Figure \ref{fsl_explain} illustrates an episode of a metric based FSL algorithm. This consists of two stages: a) embedding and b) classification. After their definition, the support and query set are both embedded in a feature space using a neural network, $F(image, \vec{\theta})$, where $\vec{\theta}$ are the parameters of the network. Once the support and query set are embedded, the resulting feature vectors of the query set are compared to the feature vectors of the support set, using a specific distance metric, $D(\vec{x_i},\vec{x_j})$, where $\vec{x_n} \in \mathbb{R}^{s}$, $s \in \mathbb{N}$. The query samples are then assigned to the classes of the support vectors that are closest within the vector space, according to metric D. In our example case of Si(001):H, the support and query sets consist of 2.5 nm × 2.5 nm crops of individual defects taken from the STM image in Figure \ref{explanation}(a), where class (i) is double dangling bonds \cite{doubleDBs}, class (ii) is siloxane \cite{Si(001)-H_features}, and class (iii) is single dangling bonds \cite{Si(001)-H_features}. If these 2D STM crops are flattened, then they can also be viewed as 1D vectors. In essence, the aim is to train a network that can perform a dimensionality reduction on these vectors to remove redundancy while retaining any important information about the image crops, thus allowing for a better comparison between query and support set.

As illustrated in Figure \ref{workflow}(e) and (f), before passing all feature crops to the FSL module for classification, the user is first prompted to assess a subset of the data and define a support set of K examples of N feature classes, thus initializing the FSL module. All remaining crops are then fed directly to the FSL module as the query set. This support set can be redefined by the user, as required. Once all defects from an image have been classified by the FSL module, this result is combined with the U-Net binary segmentation to produce a fully segmented mask, as shown in Figure \ref{workflow}(h). This final output includes pixel coordinates of all defects.

As discussed below, we test the four different FSL algorithms summarized in Table \ref{tab: table1}: standard implementations of prototypical, matching, and simple-shot algorithms \cite{prototypical_net} \cite{matching_net} \cite{simpleshot_net}, and a modified version of the relation network \cite{relation_net} supporting a variable number of classes N (details provided in the Appendix \ref{relnet_appendix}). Due to its use of a second neural network for a distance metric, the relation network has a larger number of parameters and so requires extended training for accuracy to converge. This should be considered when comparing it to the other models.

To standardise our networks, and provide an even comparison, all network backbones use the same conv4 architecture that has been widely used throughout the ML community for accurate embeddings \cite{conv4_benchmark1} \cite{resnet_conv4}. The only augmentation used in training of the FSL networks is rotation of the defects at angles of 0, 90, 180, 270 degrees.

\begin{table*}
    \begin{center}
        \begin{tabular}{||l | l | l | l || }
        \hline
         FSL network &  Training type & Training length & Distance metric \\ 
        \hline
         Protoypical \cite{prototypical_net} &  Episodic & 2000 episodes & Euclidean \\ 
         Matching \cite{matching_net} &   Episodic & 2000 episodes & Cosine \\
         Relation \cite{relation_net} &  Episodic & 5000 episodes & Multi-layer perceptron (MLP) \\
         Simple shot \cite{simpleshot_net} &  Normal & 20 epochs & Euclidean \\
         \hline
        \end{tabular} 
    \caption{Different FSL networks used. The training length was chosen such that the test accuracy converged. The relation network required more episodes since it has more weights to optimise than the other networks.}
    \label{tab: table1}
    \end{center}
\end{table*}

\subsection{Workflow Benefits}

The image recognition framework presented here is well-suited to STM data analysis because it relaxes the requirement for labelled data. This relaxation not only unburdens the user but also introduces a flexibility that can address the variability and the overwhelming detail inherent in atomic resolution STM data. For example, the shape of the STM tip may change during measurement, having an influence on the appearance of features in the images. If a human user can recognize the features across tip changes, then the images can still be accurately segmented with our approach by adjusting the FSL support set, on the fly, in step (iv). This on-the-fly adjustment of the support set can also be used to tag anomalous features, marking them for exclusion from final statistics. \highlight{For example, the two anomalies seen in Figure~\ref{workflow}(h) are at least twice as large as the other defects in the image and clearly different.} This is an advance on the supervised learning of Rashidi et.al. \cite{Rashidi_2020} where no provision is made for anomalies, meaning they are incorrectly classified as one of the predefined classes. While the unsupervised clustering technique of Ziatdinov et al. \cite{ziatdinov2020} can deal with anomalies, no accuracies are provided since the approach is unsupervised. \highlight{Our workflow allows users to explicitly exclude anomalous or non-representative features, enabling the FSL network to focus on recurring defect types of interest rather than being confounded by rare events. Although the use of neural networks in place of hand-engineered feature vectors reduces direct interpretability, it enables the learning of richer, more expressive representations that may capture subtle structural variations which may be difficult to encode manually. This data-driven representation learning also improves robustness to noise, and reduces the need for problem-specific feature design, making the approach more easily transferable across different substrates and imaging conditions.}
Furthermore, while the workflow presented here is a specific implementation of our image recognition framework, our modular code is flexible, permitting alternative implementations. \del{For example, the FSL network might be used to perform a dimensionality reduction of the data, followed by unsupervised clustering on these latent space vectors, as done by Ziatdinov et al.}

\section{Results \& Discussion}
We now assess the workflow when used to segment STM images from each of our three test cases: Si, Ge, and TiO$_2$. For segmentation tasks, accuracies alone can be misleading. We therefore provide a second metric: the dice scores (DS), where $DS = \frac{A \cap B}{|A|+|B|}$, and A and B are any two sets. We provide both the DS for the lattice defects alone (i.e. the class of immediate interest in our test cases), and also the total DS for both the background lattice \textit{plus} defects, thus allowing direct comparison with the DSs provided by Rashidi et al. \cite{Rashidi_2020}.

\subsection{U-Net Segmentation}

Table \ref{tab: UNet_acc} summarizes the U-Net performance on the three test cases. Here we achieve accuracies that are all above 90\%, with the U-Net performing best on the germanium and worst on the titania. With regards to the dice score metric, we find the network also performs best on the germanium, with a DS = 0.99. Within our workflow, the FSL network classifies the defects, therefore, the U-Net need only provide the FSL module with an accurate estimate of the defects and not the background lattice. It is the DS for the defects alone then, that is most important to gauge the effectiveness of the U-Nets. A high accuracy but low DS, as in the case of the silicon (96.3\% and 0.64  respectively), suggests the network struggles with boundary precision. The network likely struggles more with boundaries of the small surface defects in the case of the silicon data, since this data has the highest nm to pixel ratio. 

As noted, Rashidi et al. provide an overall dice score including defects plus background and report a total DS = 0.86 for their supervised segmentation of the Si(001):H surface \cite{Rashidi_2020}. This is significantly lower than our result of total DS = 0.96, but it should be remembered that their network assessed seven classes. This is inherently harder than the two classes used in our case but also comes at the cost of requiring five times more labelled data at this segmentation stage.

\begin{table}[!htb]
\footnotesize
    \begin{center}
        \begin{tabular}[width=5cm]{|| c | c | c | c ||}
        \hline
         Surface & Val acc & Dice score  & Dice  \\
         & &  (defects) & score (total) \\
        \hline
         Si(001)& \multirow{2}{*}{96.3\%} & \multirow{2}{*}{0.64} & \multirow{2}{*}{0.96} \\ 
         /Si(001):H &&& \\
         Ge(001)    &   99.0\% & 0.87  & 0.99 \\
         TiO$_2$(110)     &  90.1\%  & 0.88 & 0.96 \\
         \hline
        \end{tabular} 
    \caption{ Validation accuracies and dice scores of the U-Net for the three different surfaces.} 
    \label{tab: UNet_acc}
    \end{center}
\end{table}

\subsection{FSL Classifications}

Alongside the four FSL networks explored, we also include a k-nearest neighbour (kNN) classification on the bare pixels \highlight{(k=1, and k=3)}. This provides a baseline for a comparison of each of the embedding methods. Mean shift clustering was also investigated but found to be very sensitive to the bandwidth (i.e. the radius used to calculate the mean in each iteration step), thus proving too difficult to automate.

To assess the suitability of our FSL module for analysis of STM data corresponding to a known substrate exposed to an unknown adsorbate, we first test the FSL algorithms when classifying unseen features in the silicon data after having been trained on a subset of known features on this same surface. Although the Si(001):H surface contains many well understood defect types \cite{Si(001)-H_features}, the majority of these features are relatively rare and not easily included in the training data. By grouping the bare Si(001), Si(001):H and Si(001):H:AsH$_3$ surfaces together, we provide sufficient, regularly occurring defects within the dataset to allow for training, validation, and test data that do not share classes. In this way the network is trained on the known substrate, silicon, and tested on a set of unseen features that are a stand-in for a new unknown adsorbate.

In contrast, the germanium and titania cases are introduced as examples where both substrate and adsorbate are unknown. We test the FSL algorithms on the Ge(001):AsH$_3$ after having been trained on only the silicon and titania data and on the TiO$_2$(110) after having only been trained on silicon and germanium data. That is, for the germanium and titania examples, the training and validation datasets do not contain any germanium and titania data respectively, but the test sets do. The class splits for the training, testing, and validation sets in all of the test cases are detailed in Table \ref{tab:acc}. Note, in the case of kNN the training data appears identical to the test data due to the terminology for kNN being different to that of the FSL algorithms, which all contain neural networks. Full details of every feature type used in training, test and validation are provided in Appendix \ref{defects_appendix}.

\begin{table*}[!htb]
\footnotesize
    \begin{center}
        \begin{tabular}{|| c | c | c | c | c | c || }         
        \hline
     \multicolumn{6}{ |c| }{Si(001):H:AsH$_3$ (test set 
 of classes 5-8) } \\
         \hline
         \multirow{2}{1em}{Model} & Training & Acc &  Acc & \highlight{Prec} & \highlight{Recall} \\ 
          & \& validation data & (4-way, 1-shot) & (4-way, 3-shot) & (4-way, 3-shot)& (4-way, 3-shot)  \\
        \hline
         Prototypical &Classes $\pm$1-4, $\pm$9-15&  \textbf{ 93.6$\pm$ 0.9\%} &  \textbf{ 98.9$\pm$0.3\%} & \textbf{0.99}&\textbf{0.99}\\ 
         Matching  &Classes $\pm$1-4, $\pm$9-15&  84.3$\pm$1.4  \% & 94.5 $\pm$ 0.9\%& 0.95 & 0.94 \\
         Relation &Classes $\pm$1-4, $\pm$9-15&  54.7$\pm$2.0 \%& 58.6$\pm$1.7\%& 0.52& 0.59\\
         Simple shot (conv4) &Classes $\pm$1-4, $\pm$9-15&  88.5$\pm$1.7\%&   94.7$\pm$0.7 \%& 0.95 & 0.95\\
         Simple shot (ResNet18) & ImageNet  &   77.8$\pm$ 1.6\%&   87.7$\pm$0.8 \%& 0.88 & 0.88\\
         kNN (k=1) on bare pixels & Classes 5-8& 76.4$\pm$ 2.1\%& 90.5$\pm$0.9\% & 0.92 & 0.90 \\      
         \hline
     \multicolumn{6}{ |c| }{Ge(001):AsH$_3$ (test set of classes 9-13)} \\
         \hline
         \multirow{2}{1em}{Model} &Training &Acc &  Acc &\highlight{Prec} & \highlight{Recall}  \\
         & \& validation data & (4-way, 1-shot) &(4-way, 3-shot) & (4-way, 3-shot)& (4-way, 3-shot) \\
         \hline
           Prototypical & Classes $\pm$1-8, $\pm$14-15 & \textbf{54.5$\pm$2.0\%}& \textbf{63.4$\pm$2.1\%} & \textbf{0.62}&\textbf{0.62}\\
         Matching     & Classes $\pm$1-8, $\pm$14-15 & 50.1$\pm$2.1\% & 57.0$\pm$ 2.2\%& 0.57 & 0.56\\
         Relation     & Classes $\pm$1-8, $\pm$14-15 &  38.9$\pm$1.7 \%&  38.5$\pm$1.6\%& 0.54&0.55\\
         Simple shot (conv4) & Classes $\pm$1-8, $\pm$14-15&  52.4$\pm$2.0 \%&  60.3$\pm$2.1\% & 0.60 & 0.60\\
         Simple shot (ResNet18) & ImageNet &   45.8$\pm$ 1.9\%&   54.9$\pm$1.7\%&0.54&0.54\\
         kNN (k=1) on bare pixels & Classes 9-13& 47.1$\pm$ 1.5\%&  62.4$\pm$ 1.7\% & 0.62&0.63\\
         \hline
     \multicolumn{6}{ |c| }{TiO$_2$(110) (test set of classes 14-15)} \\
        \hline
         \multirow{2}{1em}{Model} &Training &Acc &  Acc & \highlight{Prec} & \highlight{Recall} \\
         & \& validation data & (2-way, 1-shot) & (2-way, 3-shot) & (2-way, 3-shot) & (2-way, 3-shot) \\
        \hline
         Prototypical & Classes $\pm$1-13 & 66.4$\pm$2.3 \% & 72.5$\pm$2.1\% & 0.75 & 0.73\\ 
         Matching     & Classes $\pm$1-13&  63.8$\pm$2.5\%& 66.7 $\pm$2.6 \%&0.73&0.67\\
         Relation     & Classes $\pm$1-13&  62.6$\pm$3.3 \%& 53.5 $\pm$ 1.5\%&0.54&0.54\\
         Simple shot (conv4)  & Classes $\pm$1-13&  69.2$\pm$3.1\%&  75.4$\pm$3.0\% & 0.76 & 0.76\\
         Simple shot (ResNet18) & ImageNet& \textbf{75.9$\pm$ 2.4\%}&  \textbf{85.2$\pm$2.0\%}&\textbf{0.86}&\textbf{0.85}\\
         kNN (k=1) on bare pixels & Classes 14-15&  57.5$\pm$2.9\%&  71.7$\pm$2.6\%&0.72&0.72\\
         \hline
        \end{tabular} 
\caption{Accuracies,macro-averaged precisions and macro-averaged recalls for the different FSL algorithms on the silicon, germanium, and titanium dioxide datasets. Classes 1-8 are from Si(001) and Si(001):H:AsH$_3$, classes 9-13 are from Ge(001):AsH$_3$, and classes 14-15 are from TiO$_2$(110). Negative classes indicate the inverse class (e.g. -1 is the inverse of class 1). Results shown here are a subset of all results obtained and are the best outcomes for each algorithm-dataset-substrate triplet. Results from the best performing algorithm for each surface are highlighted in bold. More details about the classes are provided in Appendix \ref{defects_appendix}. More details on the inverse classes in the ablation study are presented in section \ref{ablation_study}. Conv4 and ResNet18 refer to the backbones used for the simple shot network. Accuracies are averaged over 100 episodes and with 95\% confidence interval, with query sets containing 15 images per class. Precision and recall are calculated cumulatively over all episodes to better illustrate how the network deals with difficult classes in different scenarios.} 
    \label{tab:acc}
    \end{center}
\end{table*}

\begin{figure}[!htb]
    \centering
    \includegraphics[width=\linewidth]{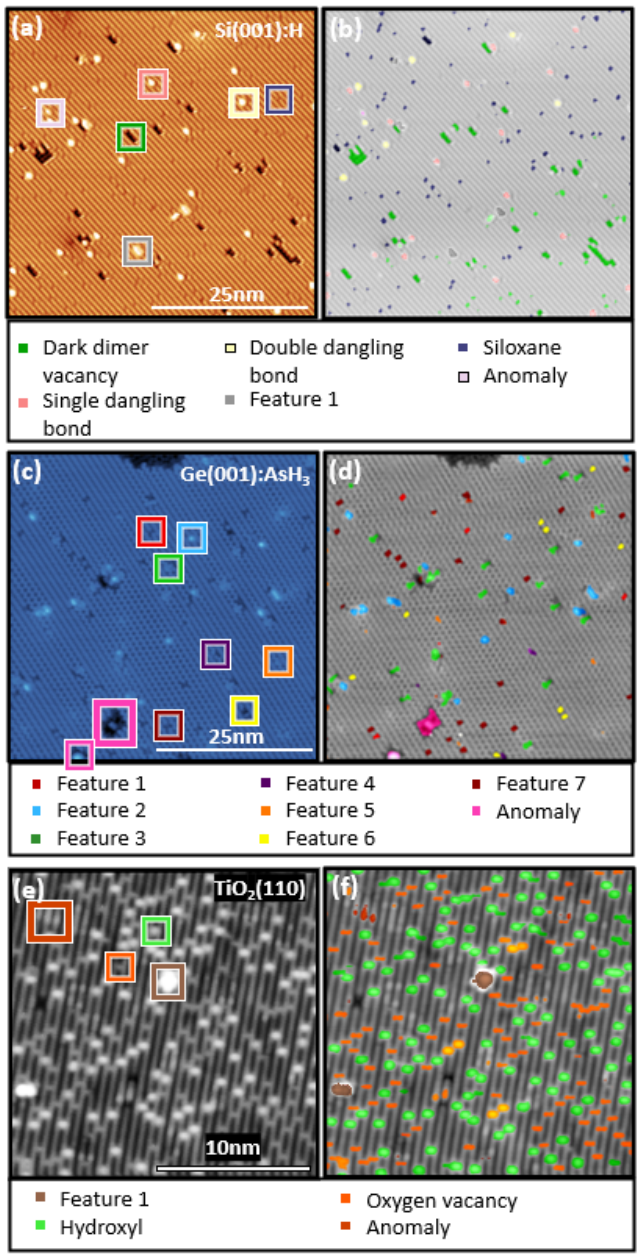}
    \caption{Final image segmentation plus feature classification on the three substrates: (a) Si(001) (b) Ge(001):AsH$_3$ (c) TiO$_2$(110). The colour coded squares around the features are the ones that were given as examples to the few-shot learning algorithm. In each case we use the best performing algorithm for that substrate from Table \ref{tab:acc}, where the corresponding accuracies can be found.} 
    \label{final_seg}
\end{figure}

Alongside the training and validation details of our test cases, Table \ref{tab:acc} also presents accuracies for 1- and 3-shot FSL classifications performed by each of the FSL algorithms \highlight{as well as the macro-averaged precisions and recalls for the 3-shot case. We provide confidence intervals due to the variation from the choice of support and query sets during testing. However, there are two additional, significant sources of randomness that we have also considered while training our models: the random initialisation of the model weights, and the data split between the training and validation sets. We therefore include data exploring the variability of the accuracy due to these sources in Appendix~\ref{randomnedd_appendix}.} \del{Comparing the results from each embedding across the different test cases, we find the prototypical and algorithms tend to produce the highest accuracy. Even with as little as one labelled image of the features, these algorithms can achieve a 94\% accuracy when applied to the silicon data. Next best is the simple shot model, which we find performs much better when using conv4 instead of ResNet18 for the embedding. We also note that computation was much slower when using Resnet18, due to the large size of this network.  Finally, the relation network performs drastically worse than all other networks, specifically in the 1-shot classification. The distance function, D($\vec{x_i}$, $\vec{x_j}$), of the relation network is a second neural network, whereas the others use fixed distance functions, therefore, it is likely that low accuracies of the relation network are due in part to over fitting to the training data. Nevertheless, the embeddings of nearly all the FSL networks examined offer an increase in accuracy compared to the kNN baseline, without further training. This is the case even in the most challenging example of the Ge(001):AsH$_3$ data where the accuracies are considerably lower due to subtler differences between detected features. The results from the best performing FSL algorithm for each test case are presented in Figure \ref{final_seg}, here a segmented and fully classified image is presented for each studied surface.}

\highlight{Comparing the performance of each embedding across the different test cases, we find that the prototypical and simple shot algorithms consistently achieve the highest accuracies among the neural-network-based methods. Notably, even with only a single labelled example per class, the prototypical network attains an accuracy of approximately 94\% on the silicon dataset, with the simple shot model providing the next best performance.

For two of the three substrates studied, the simple shot model trained on substrate-specific data outperforms the ResNet18-based approach, despite the latter’s substantially larger network capacity and access to a much larger training dataset. In addition, inference using ResNet18 is considerably slower due to its increased computational complexity. In contrast, the relation network performs significantly worse than all other evaluated models. Unlike the other methods, which rely on fixed distance metrics, the relation network employs a learned distance function D($\vec{x_i}$, $\vec{x_j}$) implemented as a second neural network. Given the limited size of the training dataset, this additional model complexity likely leads to overfitting, contributing to its poor performance.

Nevertheless, with the exception of the relation network, the embeddings produced by the few-shot learning (FSL) models examined consistently outperform the kNN baseline in the 1-shot setting, even without additional training. This improvement persists in the most challenging case, Ge(001):AsH$_3$, where classification accuracies are generally lower due to more subtle distinctions between detected features. However, the 3-shot setting is arguably of greater practical relevance, as acquiring two additional labelled examples is relatively straightforward. In this regime, the advantage of the FSL methods is only maintained for the silicon dataset, where the training and validation sets include defects from the same substrate as the test set.

The results from the best-performing FSL algorithm for each test case are presented in Figure~\ref{final_seg}, which shows fully segmented and classified images for each of the surfaces studied.}

\subsection{Improving Accuracies: Ablation Study} \label{ablation_study}
\label{section:ablation}

Finally, is it possible to improve on the reported accuracies without resorting to labelling more data? The classification accuracies of the FSL networks depend on how well they can embed the images into the latent vector space. Intuition would suggest that a larger variety of defect classes seen during training would result in higher quality embedding. To test this, we supplement the datasets with artificial defects, formed by inverting the pixel values of real defect images to produce negative images, and in effect, new defect types. In this way, we show that the data supports our intuition in some cases; the accuracies of our models increase when we increase the number of classes in $\mathscr{C}(T_{tr})$. For example, when using the Ge(001):AsH$_3$ data as our validation set, we double the size of our training and test set from classes 1-8, and 14-15 to classes 1-8, and 14-15 plus all inverses. As a result, \del{the prototypical network accuracy increases by 5\% in the 1-shot case, while}the matching network improves by 12\%. \highlight{In the case of silicon, we see an increase of 5\% for the 3-shot prototypical network when we artificially increase our dataset in this way. This initial success is promising and establishes a path to further increase the effectiveness of our STM segmentation approach without increasing the demand on manual data labelling. In the appendix \ref{ablation_study_appendix} we provide a table with further exploration of accuracy improvements through artificial dataset size increases.}

\section{Conclusion}
We have developed a machine-learning-based image-recognition framework tailored specifically for the segmentation of atomically resolved STM images involving previously unseen adsorbates and substrates. \highlight{Compared with earlier approaches - such as that of Hellerstedt et al. \cite{hellerstedt2022counting} and other data-intensive ML pipelines \cite{zhu2022deep, wei2025two} - our method offers greater flexibility: anomalous structures can be readily excluded (in contrast to the data-intensive ML studies), and class definitions can be modified with minimal effort while requiring far fewer labelled examples (unlike the two ML-based approaches). Even without extensive hyperparameter tuning or architectural optimisation,} the models achieve \del{1}\highlight{3}-shot accuracies of up to \del{93\%}\highlight{99\%}, \del{62\%}\highlight{62\%}, and \del{70\%}\highlight{85\%} for our silicon, germanium, and titania test cases, respectively. \highlight{By incorporating some of the more complex architectures for feature extracting seen in previous studies, such as that by Wei et al., we believe that the accuracy may further improve \cite{wei2025two}.}

While an accuracy of 99\% is sufficient for drawing confident conclusions about surface chemistry, a \del{62\%}\highlight{62\%} rate is not. Nevertheless, the adaptability and low data requirements of the proposed framework make it a practical tool \highlight{for rapid image analysis and annotation, at the point of data acquisition, within laboratory workflows}, where even intermediate accuracies (e.g., around 70\%) can provide rapid and actionable insight into ongoing experiments. 

The bottleneck to this approach is the need for new segmented labelled data when working with a new surface. We have therefore introduced a pipeline to ease the segmentation by implementing unsupervised methods in a training data labelling module. Next, a coarse segmentation module discerns lattice irregularities, demonstrating robust performance even with as few as 4 or 5 scans. Finally, a classification module uses FSL networks to embed the surface features into a latent space before classifying them. The FSL networks tested here exhibit optimal results with data from the same surface used for training, but in all test cases, including unseen surfaces, these networks can elevate accuracies beyond traditional clustering methods on raw pixels \highlight{in the 1-shot case. However, in the 3-shot case, we see that the kNN (k=1) performs similarly if the substrate in the test set was not seen during training. } \del{, even when the application of traditional kNN shows accuracies as high as 75\% in the 1-shot case.} Furthermore, we have also presented a novel architecture for the relation network that allows for a variable number of classes without loss of accuracy. 

Finally, rather than develop optimised networks for our data, in this study we have used standard architectures shown to work well in previous studies. This approach has simplified the initial assessment of FSL techniques for STM image segmentation, but it also leaves room for improvement. Having established the suitability of FSL for STM image recognition, improvements should now be realised by implementing a hyperparameter exploration. In this way, progress can be continued towards an efficient and material-agnostic, automatic segmentation of STM images with minimal labelled data requirements.

\section{Experimental}
The three substrate systems (Si(001):H:AsH$_3$, Ge(001):AsH$_3$, TiO$_2$) are prepared as detailed in references \cite{AsH3_on_Si}, \cite{GeAsH3}, and \cite{TiO2_prep} respectively. The silicon systems were measured using an Omicron variable temperature STM, while the germanium was measured using an Omicron low temperature STM. Silicon and germanium were all measured at room temperature. The titania with an Omicron low temperature STM at 78 K. 
Training of the U-Net was carried out over 60 epochs, with a batch size of 500, a learning rate of 0.01 that decays by 0.1 every 15 epochs, and a binary cross entropy loss. Training of the FSL networks were carried out with a learning rate of 0.001 (their training lengths were given in Table \ref{tab: table1}) with a cross entropy loss. Simple shot used batch sizes of 32, while the rest were trained using episodes.

\section{Acknowledgments}
This project was financially supported by the Engineering and Physical Sciences Research Council (EPSRC) [grant numbers EP/V027700/1, and EP/W000520/1], Innovate UK [grant number UKRI/75574], and ERC AdvG ENERGYSURF. N.L.K., M.T and E.V.S.H were partly supported by the EPSRC Centre for Doctoral Training in Advanced Characterisation of Materials [grant number EP/L015277/1] and Nanolayers Research Computing. We thank Emily V.S Hofmann for providing us with the Ge(001):AsH$_3$ data in this study.

\section{Conflict of Interest}
The authors declare no conflict of interest.

\section{Data Availability Statement}
The data that support the ﬁndings of this study are openly available at https://doi.org/10.5281/zenodo.15525935.
\section{Code Availability}
The code and models trained can be found on github at https://github.com/nickkolev97/FSL\_STM.

\bibliography{bibliography}

@Article{Rashidi_2020,
  author    = {Mohammad Rashidi and Jeremiah Croshaw and Kieran Mastel and Marcus Tamura and Hedieh Hosseinzadeh and Robert A Wolkow},
  title     = {Deep learning-guided surface characterization for autonomous hydrogen lithography},
  year      = {2020},
  number    = {2},
  pages     = {025001},
  volume    = {1},
  doi       = {10.1088/2632-2153/ab6d5e},
  journal  = {Machine Learning: Science and Technology},
  publisher = {IOP Publishing},
}

@article{rashidi_tip,
  title={Autonomous scanning probe microscopy in situ tip conditioning through machine learning},
  author={Rashidi, Mohammad and Wolkow, Robert A},
  journal={ACS Nano},
  volume={12},
  number={6},
  pages={5185--5189},
  year={2018},
  publisher={ACS Publications}
}

@Article{OGordon_tip,
  author   = {Gordon, O. and D’Hondt, P. and Knijff, L. and Freeney, S. E. and Junqueira, F. and Moriarty, P. and Swart, I.},
  journal  = {Rev. Sci. Instrum.},
  title    = {{Scanning tunneling state recognition with multi-class neural network ensembles}},
  year     = {2019},
  issn     = {0034-6748},
  number   = {10},
  pages    = {103704},
  volume   = {90},
  doi      = {10.1063/1.5099590},
  fjournal = {Review of Scientific Instruments},
}

@Article{gordon2020embedding,
  author    = {Gordon, Oliver M and Junqueira, Filipe LQ and Moriarty, Philip J},
  journal   = {Machine Learning: Science and Technology},
  title     = {Embedding human heuristics in machine-learning-enabled probe microscopy},
  year      = {2020},
  number    = {1},
  pages     = {015001},
  volume    = {1},
  publisher = {IOP Publishing},
}

@Article{deepspm,
  author  = {Krull, A. and Hirsch, P. and Rother, C. and Schiffrin, A. and Krull C.},
  journal = {Nature},
  title   = {Artificial-intelligence-driven scanning probe microscopy},
  year    = {2020},
  issn    = {2399-3650},
  doi     = {10.1038/s42005-020-0317-3},
}

@Article{STM_DFT_simulated_scans,
  author   = {Choudhary, K. and Kevin F., Camp C. and Kalinin Sergei V. and Vasudevan, R. and Ziatdinov, M. and Tavazza, F.},
  title    = {Computational scanning tunneling microscope image database},
  year     = {2021},
  journal = {Scientific Data},
}

@Article{atomAI,
  author    = {Ziatdinov, Maxim and Ghosh, Ayana and Wong, Chun Yin and Kalinin, Sergei V.},
  journal   = {Nature Machine Intelligence},
  title     = {AtomAI framework for deep learning analysis of image and spectroscopy data in electron and scanning probe microscopy},
  year      = {2022},
  issn      = {2522-5839},
  number    = {12},
  pages     = {1101–1112},
  volume    = {4},
  doi       = {10.1038/s42256-022-00555-8},
  publisher = {Springer Science and Business Media LLC},
}

@Article{STM_noiseremoval,
  author    = {Joucken, Fr\'ed\'eric and Davenport, John L. and Ge, Zhehao and Quezada-Lopez, Eberth A. and Taniguchi, Takashi and Watanabe, Kenji and Velasco, Jairo and Lagoute, J\'er\^ome and Kaindl, Robert A.},
  journal   = {Physical Review Materials},
  title     = {Denoising scanning tunneling microscopy images of graphene with supervised machine learning},
  year      = {2022},
  pages     = {123802},
  volume    = {6},
  doi       = {10.1103/PhysRevMaterials.6.123802},
  issue     = {12},
  numpages  = {11},
  publisher = {American Physical Society},
}

@Article{STM_cycleGan,
  author    = {Abid Khan and Chia-Hao Lee and Pinshane Y. Huang and Bryan K. Clark},
  title     = {Leveraging generative adversarial networks to create realistic scanning transmission electron microscopy images},
  year      = {2023},
  number    = {1},
  volume    = {9},
  doi       = {10.1038/s41524-023-01042-3},
  journal  = {npj Computational Materials},
  publisher = {Springer Science and Business Media {LLC}},
}

@article{FSL_review,
  title={A comprehensive survey of few-shot learning: Evolution, applications, challenges, and opportunities},
  author={Song, Yisheng and Wang, Ting and Cai, Puyu and Mondal, Subrota K and Sahoo, Jyoti Prakash},
  journal={ACM Computing Surveys},
  volume={55},
  number={13s},
  year={2023},
  publisher={ACM New York, NY}
}

@inproceedings{heggan,
  title={Metaaudio: A few-shot audio classification benchmark},
  author={Heggan, Calum and Budgett, Sam and Hospedales, Timothy and Yaghoobi, Mehrdad},
  booktitle={International Conference on Artificial Neural Networks},
  pages={219--230},
  year={2022},
  organization={Springer}
}

@article{lyding,
  title={Nanoscale patterning and oxidation of H-passivated Si (100)-2$\times$ 1 surfaces with an ultrahigh vacuum scanning tunneling microscope},
  author={Lyding, JW and Shen, T-C and Hubacek, JS and Tucker, JR and Abeln, GC},
  journal={Applied Physics Letters},
  volume={64},
  number={15},
  pages={2010--2012},
  year={1994},
  publisher={American Institute of Physics}
}

@Article{NIST_SETs,
  author  = {Wang, Xiqiao and Wyrick, Jonathan and Kashid, Ranjit V. and Namboodiri, Pradeep and Schmucker, Scott W.and Murphy, Andrew and Stewart, M. D. and Silver, Richard M.},
  journal = {Communications Physics},
  title   = {Atomic-scale control of tunneling in donor-based devices},
  year    = {2020},
  volume  = {3},
  doi     = {10.1038/s42005-020-0343-1},
}

@Article{2_qubit_gate_P,
  author   = {Broome, M. A.and Gorman, S. K. and House, M. G. and Hile, S. J. and Keizer, J. G. and Keith, D. and Hill C. D. and Watson, T. F. and Baker, W. J. and Hollenberg, L. C. L. and Simmons, M. Y.},
  journal  = {Nature Communications},
  title    = {Two-electron spin correlations in precision placed donors in silicon},
  year     = {2018},
  volume   = {9},
  doi      = {10.1038/s41467-018-02982-x},
}

@Article{radar_fsl,
  author    = {Diyang Liu and Xunzhang Gao and Qinmu Shen},
  journal   = {Journal of Physics: Conference Series},
  title     = {Prototypical Network for Radar Image Recognition with Few Samples},
  year      = {2020},
  number    = {1},
  pages     = {012116},
  volume    = {1634},
  doi       = {10.1088/1742-6596/1634/1/012116},
  publisher = {IOP Publishing},
}

@Article{galaxy_FSL1,
  author    = {Zhang, Zhirui and Zou, Zhiqiang and Li, Nan and Chen, Yanli},
  journal  = {Research in Astronomy and Astrophysics},
  title     = {Classifying galaxy morphologies with few-shot learning},
  year      = {2022},
  number    = {5},
  pages     = {055002},
  volume    = {22},
  publisher = {IOP Publishing},
}

@inproceedings{proto_clip,
  title={Proto-clip: Vision-language prototypical network for few-shot learning},
  author={Jaykumar P, Jishnu and Palanisamy, Kamalesh and Chao, Yu-Wei and Du, Xinya and Xiang, Yu},
  booktitle={2024 IEEE/RSJ International Conference on Intelligent Robots and Systems (IROS)},
  pages={2594--2601},
  year={2024},
  organization={IEEE}
}

@inproceedings{UNet,
  title={U-net: Convolutional networks for biomedical image segmentation},
  author={Ronneberger, Olaf and Fischer, Philipp and Brox, Thomas},
  booktitle={Medical image computing and computer-assisted intervention--MICCAI 2015: 18th international conference, Munich, Germany, October 5-9, 2015, proceedings, part III 18},
  pages={234--241},
  year={2015},
  month={Nov},
  organization={Springer}
}

@article{prototypical_net,
  title={Prototypical networks for few-shot learning},
  author={Snell, Jake and Swersky, Kevin and Zemel, Richard},
  journal={Advances in neural information processing systems},
  volume={30},
  year={2017},
  month={Dec},
}

@article{matching_net,
  title={Matching networks for one shot learning},
  author={Vinyals, Oriol and Blundell, Charles and Lillicrap, Timothy and  Kavukcuoglu, Koray and Wierstra, Daan},
  journal={Advances in neural information processing systems},
  volume={29},
  year={2016},
  month={Dec},
}

@inproceedings{relation_net,
  title={Learning to compare: Relation network for few-shot learning},
  author={Sung, Flood and Yang, Yongxin and Zhang, Li and Xiang, Tao and Torr, Philip HS and Hospedales, Timothy M},
  booktitle={Proceedings of the IEEE conference on computer vision and pattern recognition},
  pages={1199--1208},
  year={2018},
}

@article{simpleshot_net,
  title={SimpleShot: Revisiting Nearest-Neighbor Classification for Few-Shot Learning},
  author={Wang, Yan and Chao, Wei-Lun and Weinberger, Kilian Q and van der Maaten, Laurens},
  journal={arXiv e-prints},
  pages={arXiv--1911},
  year={2019},
}

@Article{akers_prototyp_SEM,
  author   = {Akers, Sarah and Kautz, Elizabeth and Trevino-Gavito, Andrea and Olszta, Matthew and Matthews, Bethany E. and Wang, Le and Du, Yingge and Spurgeon, Steven R.},
  journal  = {npj Computational Materials},
  title    = {Rapid and flexible segmentation of electron microscopy data using few-shot machine learning},
  year     = {2021},
  issn     = {2057-3960},
  number   = {1},
  pages    = {187},
  volume   = {7},
  doi      = {10.1038/s41524-021-00652-z},
  fjournal = {npj Computational Materials},
}

@article{transferlearing_EBSD,
  title={Efficient few-shot machine learning for classification of EBSD patterns},
  author={Kaufmann, Kevin and Lane, Hobson and Liu, Xiao and Vecchio, Kenneth S},
  journal={Scientific Reports},
  volume={11},
  number={1},
  pages={8172},
  year={2021},
  publisher={Nature Publishing Group UK London}
}

@Article{GeAsH3,
  author   = {Hofmann, Emily V. S. and Stock, Taylor J. Z. and Warschkow, Oliver and Conybeare, Rebecca and Curson, Neil J. and Schofield, Steven R.},
  journal  = {Angewandte Chemie International Edition},
  title    = {Room Temperature Incorporation of Arsenic Atoms into the Germanium (001) Surface},
  year     = {2023},
  number   = {7},
  pages    = {e202213982},
  volume   = {62},
  doi      = {10.1002/anie.202213982},
  fjournal = {Angewandte Chemie International Edition},
}

@Article{PH3_on_Si,
  author   = {Warschkow, O. and Curson, N. J. and Schofield, S. R. and Marks, N. A. and Wilson, H. F. and Radny, M. W. and Smith, P. V. and Reusch, T. C. G. and McKenzie, D. R. and Simmons, M. Y.},
  journal  = {The Journal of Chemical Physics},
  title    = {{Reaction paths of phosphine dissociation on silicon (001)}},
  year     = {2016},
  issn     = {0021-9606},
  month    = {01},
  number   = {1},
  pages    = {014705},
  volume   = {144},
  doi      = {10.1063/1.4939124},
  fjournal = {The Journal of Chemical Physics},
}

@Article{AsH3_on_Si,
  author    = {Stock, Taylor JZ and Warschkow, Oliver and Constantinou, Procopios C and Bowler, David R and Schofield, Steven R and Curson, Neil J},
  journal   = {Advanced Materials},
  title     = {Single-Atom Control of Arsenic Incorporation in Silicon for High-Yield Artificial Lattice Fabrication},
  year      = {2024},
  number    = {24},
  pages     = {2312282},
  volume    = {36},
  fjournal  = {Advanced Materials},
  publisher = {Wiley Online Library},
}

@article{Boron_HLith1,
  title={Bipolar device fabrication using a scanning tunnelling microscope},
  author={{\v{S}}kere{\v{n}}, Tom{\'a}{\v{s}} and K{\"o}ster, Sigrun A and Douhard, Bastien and Fleischmann, Claudia and Fuhrer, Andreas},
  journal={Nature Electronics},
  volume={3},
  number={9},
  pages={524--530},
  year={2020},
  publisher={Nature Publishing Group UK London}
}

@article{BoronHlith2,
  title={Nanoscale imaging of mobile carriers and trapped charges in delta doped silicon p--n junctions},
  author={Gramse, Georg and K{\"o}lker, Alexander and {\v{S}}kere{\v{n}}, Tom{\'a}{\v{s}} and Stock, Taylor JZ and Aeppli, Gabriel and Kienberger, Ferry and Fuhrer, Andreas and Curson, Neil J},
  journal={Nature Electronics},
  volume={3},
  number={9},
  pages={531--538},
  year={2020},
  publisher={Nature Publishing Group UK London}
}

@article{ziatdinov2020,
  title={Robust multi-scale multi-feature deep learning for atomic and defect identification in Scanning Tunneling Microscopy on H-Si (100) 2x1 surface},
  author={Ziatdinov, Maxim and Fuchs, Udi and Owen, James HG and Randall, John N and Kalinin, Sergei V},
  journal={arXiv e-prints},
  pages={arXiv--2002},
  year={2020}
}

@article{TiO2_1,
  title={Direct visualization of defect-mediated dissociation of water on TiO2 (110)},
  author={Bikondoa, Oier and Pang, Chi L and Ithnin, Roslinda and Muryn, Christopher A and Onishi, Hiroshi and Thornton, Geoff},
  journal={Nature materials},
  volume={5},
  number={3},
  pages={189--192},
  year={2006},
  publisher={Nature Publishing Group UK London}
}

@article{TiO2_2,
  title={Structure of clean and adsorbate-covered single-crystal rutile TiO2 surfaces},
  author={Pang, Chi Lun and Lindsay, Robert and Thornton, Geoff},
  journal={Chemical reviews},
  volume={113},
  number={6},
  pages={3887--3948},
  year={2013},
  publisher={ACS Publications}
}

@Article{TiO2_3,
  author   = {Ulrike Diebold},
  title    = {The surface science of titanium dioxide},
  year     = {2003},
  issn     = {0167-5729},
  number   = {5},
  pages    = {53-229},
  volume   = {48},
  doi      = {10.1016/S0167-5729(02)00100-0},
  journal = {Surface Science Reports},
  }

@article{conv4_benchmark1,
  title={TorchBench: Benchmarking PyTorch with High API Surface Coverage},
  author={Hao, Yueming and Zhao, Xu and Bao, Bin and Berard, David and Constable, Will and Aziz, Adnan and Liu, Xu},
  journal={arXiv e-prints},
  pages={arXiv--2304},
  year={2023}
}

@Article{resnet_conv4,
  author        = {Kaiming He and Xiangyu Zhang and Shaoqing Ren and Jian Sun},
  journal       = {CoRR},
  title         = {Deep Residual Learning for Image Recognition},
  year          = {2015},
  volume        = {abs/1512.03385},
  archiveprefix = {arxiv},
  bibsource     = {dblp computer science bibliography, https://dblp.org},
  doi           = {10.48550/arxiv.1512.03385},
  eprint        = {1512.03385},
}

@Article{windowing_segmentation,
  author    = {Pielawski, Nicolas and Wählby, Carolina},
  journal   = {PLOS ONE},
  title     = {Introducing Hann windows for reducing edge-effects in patch-based image segmentation},
  year      = {2020},
  issn      = {1932-6203},
  number    = {3},
  pages     = {e0229839},
  volume    = {15},
  doi       = {10.1371/journal.pone.0229839},
  editor    = {Zhang, Jie},
  publisher = {Public Library of Science (PLoS)},
}

@Article{edge_effects,
  author        = {Carlo Innamorati and Tobias Ritschel and Tim Weyrich and Niloy J. Mitra},
  journal       = {CoRR},
  title         = {Learning on the Edge: Explicit Boundary Handling in CNNs},
  year          = {2018},
  volume        = {abs/1805.03106},
  archiveprefix = {arxiv},
  bibsource     = {dblp computer science bibliography, https://dblp.org},
  doi           = {10.48550/arxiv.1805.03106},
  eprint        = {1805.03106},
}

@Article{PhaseSep1,
  author    = {Song, Sun Kyu and Yeom, Han Woong},
  journal   = {Physical Review B},
  title     = {Atomistic origin of metal versus charge-density-wave phase separation in indium atomic wires on Si(111)},
  year      = {2021},
  pages     = {035420},
  volume    = {104},
  doi       = {10.1103/PhysRevB.104.035420},
  issue     = {3},
  numpages  = {5},
  publisher = {American Physical Society},
}

@Article{PhaseSep2,
  author    = {Shim, Hyungjoon and Lee, Geunseop},
  journal   = {ACS Nano},
  title     = {True First-Order Surface Phase Transition without Nanoscale Phase Separation},
  year      = {2023},
  volume    = {17},
  doi       = {10.1021/acsnano.3c02600},
  issue     = {12},
  publisher = {ACS Nano},
}

@Article{doubleDBs,
  author    = {Bowler, D. R. and Owen, J. H. G. and Miki, K. and Briggs, G. A. D.},
  journal   = {Physical Review B},
  title     = {Diffusion of paired hydrogen on Si(001)},
  year      = {1998},
  pages     = {8790--8793},
  volume    = {57},
  doi       = {10.1103/PhysRevB.57.8790},
  issue     = {15},
  numpages  = {0},
  publisher = {American Physical Society},
}

@article {SAMmicroscopy,
	author = {Archit, Anwai and Nair, Sushmita and Khalid, Nabeel and Hilt, Paul and Rajashekar, Vikas and Freitag, Marei and Gupta, Sagnik and Dengel, Andreas and Ahmed, Sheraz and Pape, Constantin},
	title = {Segment Anything for Microscopy},
	elocation-id = {2023.08.21.554208},
	year = {2023},
	doi = {10.1101/2023.08.21.554208},
	publisher = {Cold Spring Harbor Laboratory},
	URL = {https://www.biorxiv.org/content/early/2023/08/22/2023.08.21.554208},
	eprint = {https://www.biorxiv.org/content/early/2023/08/22/2023.08.21.554208.full.pdf},
	journal = {bioRxiv}
}

@Article{MicroNet,
  author    = {Stuckner, Joshua and Harder, Bryan and Smith, Timothy M},
  title     = {Microstructure segmentation with deep learning encoders pre-trained on a large microscopy dataset},
  year      = {2022},
  number    = {1},
  pages     = {200},
  volume    = {8},
  journal  = {npj Computational Materials},
  publisher = {Nature Publishing Group UK London},
}

@Article{TiO2_prep,
  author    = {Williams, Oscar Bentley Jerdmyr and Katsiev, Khabiboulakh and Baek, Byeongjin and Harrison, George and Thornton, Geoff and Idriss, Hicham},
  title     = {Direct visualization of a gold nanoparticle electron trapping effect},
  year      = {2022},
  number    = {2},
  pages     = {1034--1044},
  volume    = {144},
  journal  = {Journal of the American Chemical Society},
  publisher = {ACS Publications},
}

@Article{FCN,
  author        = {Jonathan Long and Evan Shelhamer and Trevor Darrell},
  journal       = {CoRR},
  title         = {Fully Convolutional Networks for Semantic Segmentation},
  year          = {2014},
  volume        = {abs/1411.4038},
  archiveprefix = {arxiv},
  bibsource     = {dblp computer science bibliography, https://dblp.org},
  doi           = {10.48550/arxiv.1411.4038},
  eprint        = {1411.4038},
}

@Article{Resnet101,
  author        = {Kaiming He and Xiangyu Zhang and Shaoqing Ren and Jian Sun},
  journal       = {CoRR},
  title         = {Deep Residual Learning for Image Recognition},
  year          = {2015},
  volume        = {abs/1512.03385},
  archiveprefix = {arxiv},
  bibsource     = {dblp computer science bibliography, https://dblp.org},
  doi           = {10.48550/arxiv.1512.03385},
  eprint        = {1512.03385},
}

@misc{kolev2025generativeimagerestorationsuperresolution,
      title={Generative Image Restoration and Super-Resolution using Physics-Informed Synthetic Data for Scanning Tunneling Microscopy}, 
      author={Nikola L. Kolev and Tommaso Rodani and Neil J. Curson and Taylor J. Z. Stock and Alberto Cazzaniga},
      year={2025},
      eprint={2510.25921},
      archivePrefix={arXiv},
      primaryClass={cs.CV},
      url={https://arxiv.org/abs/2510.25921}, 
}

@article{zhang2019method,
  title={A method to correct hysteresis of scanning probe microscope images based on a sinusoidal model},
  author={Zhang, Liansheng and Chen, Xiaobo and Huang, Jichao and Li, Hongli and Chen, Lijuan and Huang, Qiangxian},
  journal={Review of Scientific Instruments},
  volume={90},
  number={2},
  year={2019},
  publisher={AIP Publishing}
}

@article{Lowe2004SIFT,
  author    = {Lowe, David G.},
  title     = {Distinctive Image Features from Scale-Invariant Keypoints},
  journal   = {International Journal of Computer Vision},
  volume    = {60},
  number    = {2},
  pages     = {91--110},
  year      = {2004},
  doi       = {10.1023/B:VISI.0000029664.99615.94}
}

@article{wei2025two,
  title={Two-Stage Machine Learning Framework for Accurate Discrimination of Isomers and Very-Similar Molecules on Surfaces},
  author={Wei, Zixuan and Zhong, Qigang and Pan, Jinbo and Lim, Fang Han and Chi, Lifeng and Du, Shixuan},
  journal={Journal of the American Chemical Society},
  volume={147},
  number={39},
  pages={35232--35243},
  year={2025},
  publisher={ACS Publications}
}

@article{zhu2022deep,
  title={A Deep-Learning Framework for the Automated Recognition of Molecules in Scanning-Probe-Microscopy Images},
  author={Zhu, Zhiwen and Lu, Jiayi and Zheng, Fengru and Chen, Cheng and Lv, Yang and Jiang, Hao and Yan, Yuyi and Narita, Akimitsu and M{\"u}llen, Klaus and Wang, Xiao-Ye and others},
  journal={Angewandte Chemie International Edition},
  volume={61},
  number={49},
  pages={e202213503},
  year={2022},
  publisher={Wiley Online Library}
}

@article{yuan2023applying,
  title={Applying a Deep-Learning-Based Keypoint Detection in Analyzing Surface Nanostructures},
  author={Yuan, Shaoxuan and Zhu, Zhiwen and Lu, Jiayi and Zheng, Fengru and Jiang, Hao and Sun, Qiang},
  journal={Molecules},
  volume={28},
  number={14},
  pages={5387},
  year={2023},
  publisher={MDPI}
}

@article{hellerstedt2022counting,
  title={Counting molecules: Python based scheme for automated enumeration and categorization of molecules in scanning tunneling microscopy images},
  author={Hellerstedt, Jack and Cahl{\'\i}k, Ale{\v{s}} and {\v{S}}vec, Martin and Stetsovych, Oleksandr and Hennen, Tyler},
  journal={Software Impacts},
  volume={12},
  pages={100301},
  year={2022},
  publisher={Elsevier}
}

@inproceedings{kirillov2023segment,
  title={Segment anything},
  author={Kirillov, Alexander and Mintun, Eric and Ravi, Nikhila and Mao, Hanzi and Rolland, Chloe and Gustafson, Laura and Xiao, Tete and Whitehead, Spencer and Berg, Alexander C and Lo, Wan-Yen and others},
  booktitle={Proceedings of the IEEE/CVF international conference on computer vision},
  pages={4015--4026},
  year={2023}
}

@article{korventausta2009stm,
  title={STM simulation of molecules on ultrathin insulating overlayers using tight-binding: Au--pentacene on NaCl bilayer on Cu},
  author={Korventausta, Antti and Paavilainen, Sami and Niemi, Eeva and Nieminen, JA},
  journal={Surface Science},
  volume={603},
  number={3},
  pages={437--444},
  year={2009},
  publisher={Elsevier}
}

@article{blanco2004first,
  title={First-principles simulations of STM images: from tunneling to the contact regime},
  author={Blanco, Jose Manuel and Gonz{\'a}lez, Cesar and Jel{\'\i}nek, Pavel and Ortega, Jos{\'e} and Flores, Fernando and P{\'e}rez, Rub{\'e}n},
  journal={Physical Review B—Condensed Matter and Materials Physics},
  volume={70},
  number={8},
  pages={085405},
  year={2004},
  publisher={APS}
}

\newpage 

\onecolumn 
\section{Appendix}
\subsection{Datasets} \label{defects_appendix}
The full dataset used for this work includes data from the 4 different surfaces (Si(001), Si(001):H:AsH$_3$, Ge:AsH$_3$, and TiO$_2$) and contains two types of labelled data, as detailed in table \ref{tab: datasets}. These are: (1) fully segmented scans which have any defect in the lattice highlighted (except for step edges), and (2) scans with pixel coordinates and labels for the defects present on their surface. Due to the nature of STM, if we contrast inverse an image of a defect (i.e. turn a depression into a protrusion and vice-versa), we can produce a new 'defect'. This 'defect', is of course not seen in the STM data, but it can be used for training and validating our models. We show in section \ref{section:ablation} that this method of increasing our dataset increases the accuracy of our model. Although our model works with filled and empty state images, the TiO$_2$(110) dataset was only available as empty state images. In this case, we repeat the empty state images to make a tensor of the same shape, allowing us to use the same networks.

\begin{figure*}[!htb]
    \centering
    \includegraphics[width=13cm]{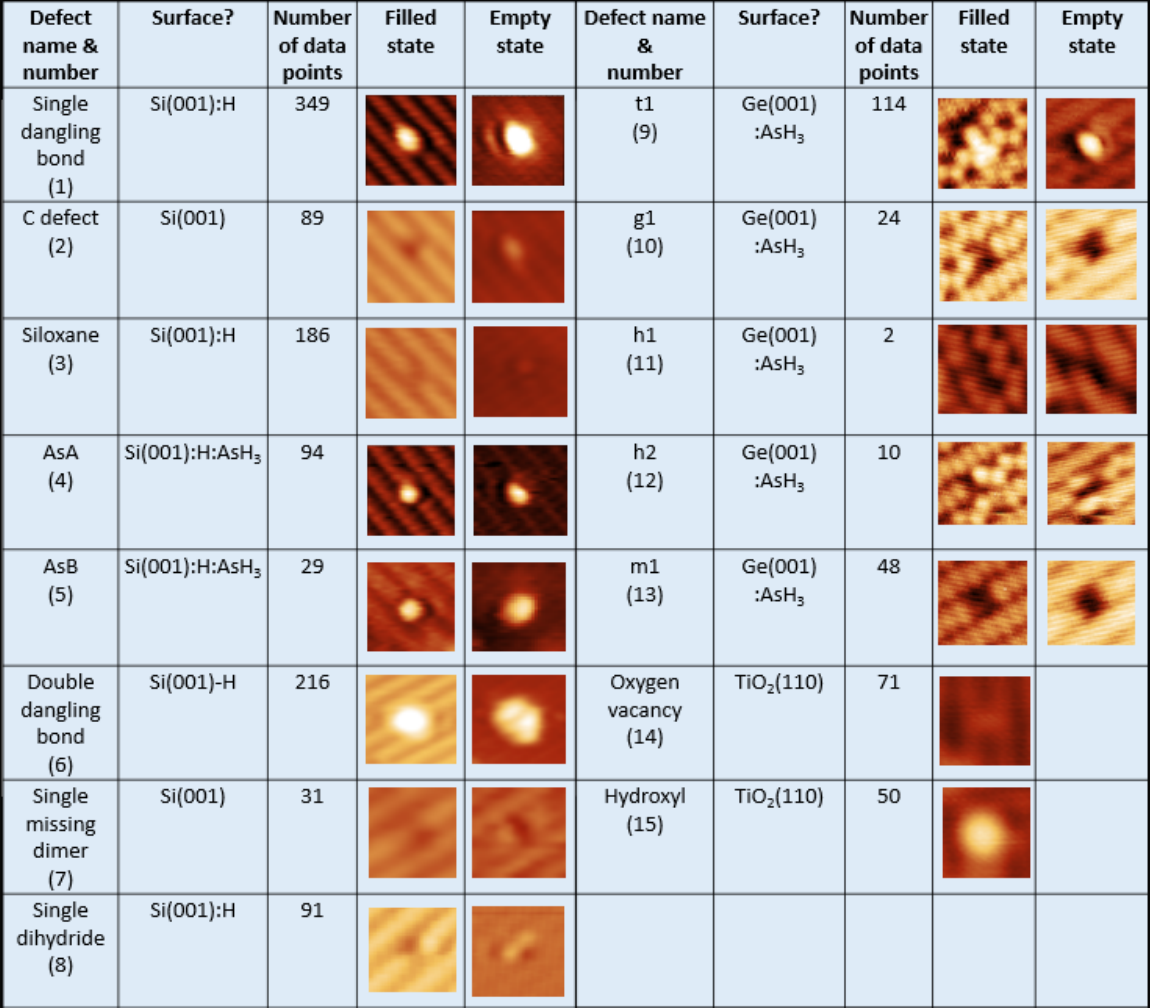}
    \caption{ Crops of the different features used to train and evaluate our networks.}
    \label{def_crops}
\end{figure*}

The different defects used for training of the FSL networks are shown in Figure \ref{def_crops}. Features h1 and h2 were very rare in the experimental data. As a result, during training, testing, and validating we augmented those crops using rotations and reflections thus producing a sufficient number of examples to allow a 4-way, 1-shot or 4-way, 3-shot episode. This is not good practice usually since the network will see augmented versions of the same data in both training and testing leading to less resilient results. However, doing so was necessary to allow us to use this data. 
\begin{table*}[!htb]
\footnotesize
    \begin{center}
        \begin{tabular}[width=13cm]{||c | c | c |  c | c | c ||}
        \hline
         Dataset & Crops (C) or & Si(001):H \&  & Ge:AsH$_3$(001) & TiO$_2$(110)  \\
         & fully segmented (FS)? &  Si(001):H:AsH$_3$  & & \\ 
        \hline
         Five 100 nm $\times$ 100 nm scans at 512 $\times$ 512 pixels & FS  & \checkmark &   $\times$ & $\times$    \\ 
         Five 50 nm $\times$ 50 nm scans at 512$\times$512 pixels & FS & $\times$ & \checkmark &$\times$  \\ 
         Four 10 nm $\times$ 10 nm scans at 512$\times$512 pixels & FS & $\times$ & $\times$  & \checkmark    \\ 
         1085 crops 11$\times$11 pixels & C & \checkmark  &  $\times$  & $\times$    \\ 
         198 crops 22$\times$22 pixels & C & $\times$ &  \checkmark  & $\times$    \\
         121 crops 61$\times$61 pixels & C & $\times$ &  $\times$  & \checkmark  \\
         \hline
        \end{tabular} 
    \caption{Description of dataset structures and data types for all data used in this study.} 
    \label{tab: datasets}
    \end{center}
\end{table*}

\clearpage

%\begin{table}[!htbp]
%\footnotesize
%    \begin{center}
%        \begin{tabular}[width=6cm]{c  c  c  c  c  c c}
%        \hline
%         Dataset & Crops (C) or & Size & Si(001):H \&  & Ge:AsH$_3$ & %TiO$_2$  \\
%         & fully segmented (FS)? &  &  Si(001):H:AsH$_3$  & & \\ 
%        \hline
%         MLSTM & C\&FS & everything &\checkmark & \checkmark & \checkmark  \\ 
%         segSTMSi & FS & 5 * 100nm$^2$ scans at 512 pixels$^2$ & \checkmark &   $\times$ & $\times$    \\ 
%         segSTMGe & FS & 5 * 50nm$^2$ scans at 512 pixels$^2$ & $\times$ & \checkmark &$\times$  \\ 
%         segSTMTiO$_2$ & FS & 4*10nm$^2$ scans at 512 pixels$^2$ & $\times$ & $\times$  & \checkmark    \\ 
%         defSTMSi & C & 1085 crops 11 pixels$^2$ & \checkmark  &  $\times$  & $\times$    \\ 
%         defSTMGe & C & 198 crops 22 pixels$^2$&$\times$  &  \checkmark  & $\times$    \\
%         defSTMTiO2 & C & 121 crops 61 pixels$^2$& $\times$ &  $\times$  & \checkmark  \\
%         \hline
%        \end{tabular} 
%    \caption{Datasets, their names, and what they contain. In general, defSTMXX contains crops of defects from one or more surfaces, and segSTM contains the fully segmented scans. If a dataset has the suffix "Inv", then it also contains the inverse of the dataset without that suffix. MLSTM is the full dataset.} 
%    \label{tab: datasets}
%    \end{center}
%\end{table}

\clearpage

\subsection{Algorithm Breakdowns}\label{algorithm_tables}

\highlight{The main text presents the algorithms through schematic illustrations. For readers who prefer a textual representation, this appendix provides the corresponding algorithms in tabular form. The tables follow the same order as the figures in the paper.}

\subsubsection{Image Recognition Workflow (Figure \ref{workflow})}

\begin{algorithm}[H]
\caption{\highlight{Image Segmentation and Few-Shot Classification}}
\begin{algorithmic}[1]
    \Require \highlight{Image $I$, number of defect classes $N$, number of human-labelled support examples per class $K$}
    \Ensure \highlight{Segmented image $S$, defect counts for each class}

    \If{\highlight{$I$ is dual-bias}}
        \State \highlight{Apply hysteresis correction to $I$}
    \Else
        \State \highlight{Duplicate single-bias image into both channels}
    \EndIf

    \State \highlight{Obtain binary defect segmentation using a U-Net model and the filled state image of I}
    \State \highlight{Extract connected components to identify individual defects}
    \State \highlight{Let $D$ be the number of detected defects}
    \State \highlight{For each defect, extract a $2.5\,\mathrm{nm} \times 2.5\,\mathrm{nm}$ crop}

    \For{\highlight{$i = 1$ to $N$}} \Comment{\highlight{For each defect class}}
        \For{\highlight{$j = 1$ to $K$}} \Comment{\highlight{Collect human-labelled support examples}}
            \State \textbf{\highlight{Human:}} \highlight{Label a defect in $I$ , $x_{i,j}$, as an exemplar of class $i$}
        \EndFor
    \EndFor

    \State \highlight{Resample all crops to 40 $\times$ 40 pixels}
    \State \highlight{Classify all remaining defects using the FSL network}
    \State \highlight{Construct final segmented image $S$ by coloring each defect according to its predicted class}

\end{algorithmic}
\end{algorithm}

\subsubsection{Unsupervised Labelling (Figure \ref{auto_segment})}

\begin{algorithm}[H]
\caption{\highlight{Unsupervised Labelling}}
\begin{algorithmic}[1]
    \Require \highlight{Image $I$, segmentation mode $M \in \{C, D\}$}
    \Ensure \highlight{Segmented image $S$, defect counts for each class}

    \State \highlight{Partition $I$ into $64 \times 64$ pixel crops with stride 32}
    \State \highlight{Apply min–max normalisation followed by z-score normalisation to each crop}

    \If{\highlight{$M = D$}}
        \State \highlight{Upscale each crop to $520 \times 520$ pixels using bilinear upsampling}
    \ElsIf{\highlight{$M = C$}}
        \State \highlight{Apply sinusoidal windowing to each crop}
        \State \highlight{Reconstruct full-resolution image from processed crops}
    \EndIf

    \State \highlight{Extract pixel-wise feature embeddings using the first two layers of a pretrained FCN–ResNet101}
    \State \highlight{Construct initial segmentation $S$ by applying $K$-means clustering with $K = 5$}

    \While{\textbf{\highlight{Human:}} \highlight{$S$ is unsatisfactory}}
        \State \textbf{\highlight{Human:}} \highlight{Specify new value of $K$}
        \State \highlight{Recompute segmentation $S$ using $K$-means with the new $K$}
    \EndWhile
    
\end{algorithmic}
\end{algorithm}

\clearpage

\subsection{Data Preprocessing} \label{app: preprocessing}

\highlight{This appendix consolidates all preprocessing steps used throughout the paper. For clarity, we organise them by module, substrate, and by whether they apply to training or inference. All modules and substrates have some preprocessing in common - all images are plane-levelled and scan line aligned. Plane levelling was performed by fitting a plane of the form $z = ax + by + c$ to the entire image, without masking or weighting. The fit was computed using the Moore-Penrose pseudo-inverse, and the estimated plane was subtracted from the image to remove the background slope. We summarise all the different crop sizes for the different substrates and modules in Table~\ref{tab: preprocessing}.}

\begin{table}[!htb]
\footnotesize
    \begin{center}
        \begin{tabular}[width=5cm]{|| c | c | c | c | c ||}
        \hline
         \multirow{2}{*}{Surface} & \multirow{2}{*}{Full image size (nm$^2$)} & Crop size for & Crop size for & Crop size for  \\
        & & Module 1 (pixels) & Module 2 (pixels) &  Module 3 (pixels) \\
        \hline
         Si(001) & \multirow{2}{*}{100 $\times$ 100} & \multirow{2}{*}{64 $\times$ 64} & \multirow{2}{*}{64 $\times$ 64} & \multirow{2}{*}{11 $\times$ 11}\\ 
         /Si(001):H &  &  & & \\
         Ge(001)   &  50 $\times$ 50 & 128 $\times$ 128  & 64 $\times$ 64 & 22 $\times$ 22 \\
         TiO$_2$(110) & 10 $\times$ 10 & N/A & 128 $\times$ 128 & 61 $\times$ 61 \\
         \hline
        \end{tabular} 
    \caption{\highlight{Summary of full image sizes (in nm$^2$) and and the crop sizes (in pixels) for all modules and substrates.}}
    \label{tab: preprocessing}
    \end{center}
\end{table}

\subsubsection{Module 1 (Unsupervised Labelling)} \label{sec: preprocessing_module1}

\highlight{Segmentation is performed using a single channel: filled state for Si(001) and Ge(001), and empty state for TiO$_2$. All images were 512 $\times$ 512 pixels; the corresponding physical dimensions are provided in Table~\ref{tab: preprocessing}. All crops were min–max normalised and z-score normalised. For detailed segmentation (i.e., to resolve defects rather than phase domains), each crop was up-sampled to 520 $\times$ 520 pixels. An exception is TiO$_2$(110), which already has a sufficiently high pixel-to-nanometre resolution.}

\subsubsection{Module 2 (U-Net)}

\highlight{Segmentation used a single channel (filled state for Si(001) and Ge(001), empty state for TiO$_2$). During training, images were cropped (crop sizes listed in Table~\ref{tab: preprocessing}), min–max normalised, and augmented. The augmentations were identical across substrates and included 90, 180, and 270 degree rotations, Gaussian blur, and simulated scan line noise (the latter two implemented similarly to \cite{kolev2025generativeimagerestorationsuperresolution}). 

During inference, the same preprocessing steps were applied except that augmentations were omitted.}

\subsubsection{Module 3 (FSL classification)}

\highlight{For FSL classification, both filled and empty state channels were used, requiring hysteresis correction on the full images prior to cropping. In the overall workflow (Figure~\ref{workflow}), hysteresis correction is performed before Module~2; however, because it is irrelevant to Module~2's U-Net segmentation, we include it here under Module~3 preprocessing. The hysteresis correction algorithm followed the method of Zhang et al.~\cite{zhang2019method}, with an additional filtering step to remove inaccurate matches produced by the SIFT algorithm~\cite{Lowe2004SIFT}. Specifically, matches were retained only if their $y$-coordinates were identical and their $x$-coordinates differed by no more than 15 pixels.

Module~3 uses small defect-centred crops (sizes given in Table~\ref{tab: preprocessing}). Each crop was normalised by subtracting its mean but not rescaled by its standard deviation, as pixel height values are essential for defect discrimination. During training, crops were augmented by 90, 180, or 270 degree rotations; no additional augmentations were applied. Finally, crops from all substrates were upsampled to 40~$\times$~40 pixels.}

\clearpage

\subsection{A quicker auto-labelling procedure} \label{knn_label_appendix}
\begin{figure*}[!htb]
    \centering
    \includegraphics[width=13cm]{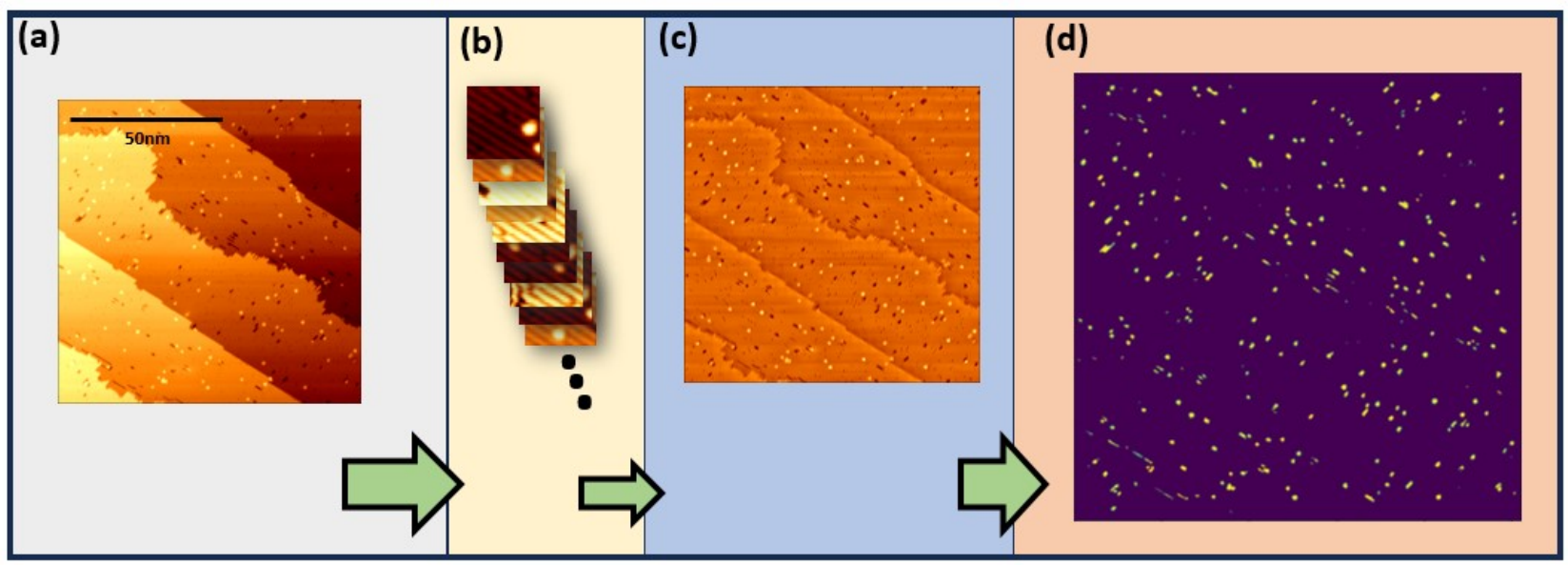}
    \caption{Less computationally heavy procedure for unsupervised labelling. The procedure from (a) to (c) is the same as that in Figure \ref{auto_segment}. From (c) to (d), we extract anything that is more or less than 3 standard deviations from the mean.}
    \label{auto_label_knn}
\end{figure*}

Figure \ref{auto_label_knn} details an automated unsupervised labelling technique that can pick out the defects without need for a GPU. Although it can pick out the defects, \del{but} it can not be tuned to segment phase structures, so is less flexible than the method presented in section 3.2. An automated unsupervised labelling technique can segment the dimers to some degree, if k-means clustering is used on Figure \ref{auto_label_knn}(c) instead of just thresholding, but not with the same accuracy as in section 3.2, especially at the step edges. The main difference between these two approaches, is that the method illustrated in Figure 7 does not perform any feature extraction using a pretrained network. Here, the k-means clustering or thresholding is performed on the pixels rather than on feature vectors.

\clearpage

\subsection{Confusion Matrices} \label{confusion_matrix_appendix}

\highlight{Since the FSL module performs a classification task, we present confusion matrices for each substrate in this section. However, special care is required when interpreting these results because FSL models are evaluated episodically. In each episode, a subset of classes is sampled and assigned temporary labels, which vary between episodes.

For the silicon and titania datasets, this does not introduce ambiguity, as these tasks involve 4-way and 2-way classification problems with exactly 4 and 2 available classes, respectively. Consequently, the same set of classes is present in every episode. In contrast, the germanium dataset comprises five classes, while evaluation is performed in a 4-way setting. As a result, not all classes are included in every episode, and the specific combination of classes varies across episodes.

To construct a meaningful confusion matrix for the germanium case, we therefore aggregate predictions across all episodes using the true class identities rather than the temporary episode-specific labels. This approach allows us to assess class-wise performance across the entire dataset and results in a $5\times5$ confusion matrix, despite the underlying task being 4-way classification.}

\begin{figure*}[!htb]
    \centering
    \includegraphics[width=15cm]{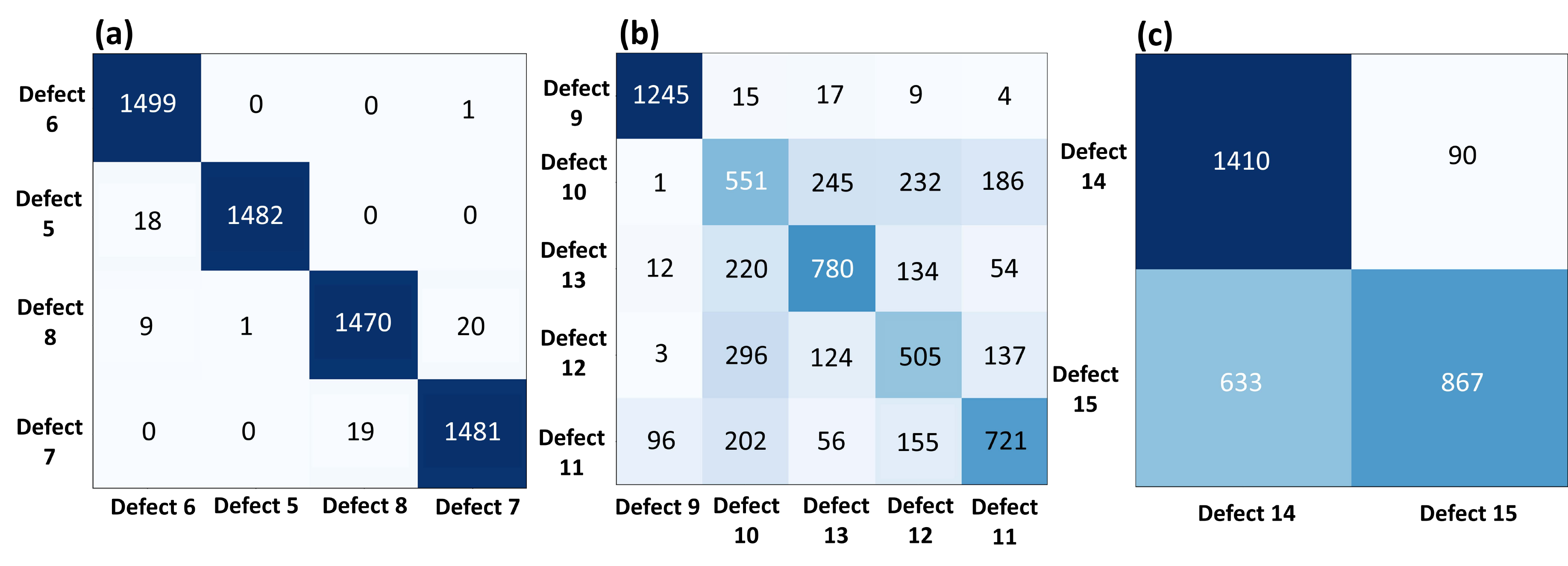}
    \caption{\highlight{Confusion matrices for the best performing 3-shot models for all three substrates: (a) Si(001):H, (b) Ge(001):AsH$_3$, and (c) TiO$_2$(110).}}
    \label{conf_mat}
\end{figure*}

\clearpage

\subsection{Ablation study continued} \label{ablation_study_appendix}
Tables \ref{tab: acc_abl_2_3_shot} and \ref{tab: acc_abl_2_1_shot} show the full array of different networks trained and their training and validation data. They illustrate the effects of training the FSL networks using a larger variety of data.

\begin{table*}[!htb]
\footnotesize
    \begin{center}
        \begin{tabular}{|| c | c | c | c | c || }         
        \hline
     \multicolumn{5}{ |c| }{Si(001):H:AsH$_3$ (test set 
 of classes 5-8) } \\
         \hline
         \multirow{2}{1em}{Model} & Training & Acc & \highlight{Prec} & \highlight{Recall}  \\ 
          & \& validation data & (4-way, 3-shot) & (4-way, 3-shot)& (4-way, 3-shot)  \\
        \hline
         Prototypical &Classes $\pm$1-4, $\pm$9-15&  \textbf{ 98.9$\pm$0.3\%} & \textbf{0.99}&\textbf{0.99}\\ 
         
         Prototypical &Classes 1-4, 9-15&94.2$\pm$0.1\% & 0.94&0.94\\ 

         Matching  &Classes $\pm$1-4, $\pm$9-15& 94.5 $\pm$ 0.9\%& 0.95 & 0.94 \\         
         
         Matching  &Classes 1-4, 9-15& 76.02 $\pm$ 1.1\%& 0.77 & 0.76 \\
         
         Relation &Classes $\pm$1-4, $\pm$9-15&  61.4 $\pm$ 1.8\%& 0.52& 0.61\\
         
         Relation &Classes 1-4, 9-15& 58.6 $\pm$ 1.7\%& 0.52& 0.59\\
         
         Simple shot (conv4) &Classes $\pm$1-4, $\pm$9-15& 94.7$\pm$0.7 \%& 0.95 & 0.95\\
         
         Simple shot (conv4) &Classes 1-4, 9-15& 85.9$\pm$1.0 \%& 0.87 & 0.86\\
         
         Simple shot (ResNet18) & ImageNet  & 87.7$\pm$0.8 \%& 0.88 & 0.88\\
         kNN (k=1) on bare pixels & Classes 5-8& 90.5$\pm$0.9\% & 0.92 & 0.90 \\ 
         kNN (k=3) on bare pixels & Classes 5-8& 82.6$\pm$1.6\% & 0.86 & 0.83 \\     
         \hline
     \multicolumn{5}{ |c| }{Ge(001):AsH$_3$ (test set of classes 9-13)} \\
         \hline
         \multirow{2}{1em}{Model} &Training & Acc & \highlight{Prec} & \highlight{Recall}  \\
         & \& validation data & (4-way, 3-shot) & (4-way, 3-shot)& (4-way, 3-shot) \\
         \hline
         Prototypical & Classes $\pm$1-8, $\pm$14-15 & \textbf{ 63.4$\pm$2.1\%} & \textbf{0.62}&\textbf{0.62}\\

         Prototypical & Classes 1-8, 14-15 & 63.4$\pm$2.1\% & 0.63& 0.63 \\
        
         Matching     & Classes $\pm$1-8, $\pm$14-15 &  57.0$\pm$ 2.2\%& 0.57 & 0.56\\

         Matching     & Classes 1-8, 14-15 & 44.9$\pm$2.1\% & 0.43 & 0.44\\

         Relation     & Classes $\pm$1-8, $\pm$14-15 & 38.5$\pm$ 1.6\%& 0.54&0.55\\
         
         Relation     & Classes 1-8, 14-15 & 37.5$\pm$ 1.5\%& 0.41&0.38\\
         
         Simple shot (conv4) & Classes $\pm$1-8, $\pm$14-15&  60.3$\pm$2.1\% & 0.60 & 0.60\\
         
         Simple shot (conv4) & Classes 1-8, 14-15& 57.4$\pm$2.3\% & 0.57 & 0.56\\
         
         Simple shot (ResNet18) & ImageNet & 54.9$\pm$1.7\%&0.54&0.54\\
         kNN (k=1) on bare pixels & Classes 9-13& 62.4$\pm$ 1.7\% & 0.62&0.62\\
         
         kNN (k=3) on bare pixels & Classes 9-13& 51.0$\pm$1.7\% & 0.56&0.51\\
         \hline
     \multicolumn{5}{ |c| }{TiO$_2$(110) (test set of classes 14-15)} \\
        \hline
         \multirow{2}{1em}{Model} &Training &Acc & \highlight{Prec} & \highlight{Recall}  \\
         & \& validation data & (2-way, 3-shot) & (2-way, 3-shot) & (2-way, 3-shot) \\
        \hline
         Prototypical & Classes $\pm$1-13 & 72.5 $\pm$2.1\% & 0.75 & 0.73\\ 
         
         Prototypical & Classes 1-13 & 70.1 $\pm$2.1\% & 0.73 & 0.71\\ 
         
         Matching     & Classes $\pm$1-13& 66.7 $\pm$2.6 \%&0.73&0.67\\
         
         Matching     & Classes 1-13& 64.50 $\pm$2.4 \%&0.72&0.65\\
         
         Relation     & Classes $\pm$1-13&  53.5 $\pm$ 1.5\%&0.54&0.54 \\
         
         Relation     & Classes 1-13& 50.9 $\pm$ 1.2\%&0.51&0.51\\
         
         Simple shot (conv4)  & Classes $\pm$1-13& 75.4$\pm$ 3.0\% & 0.76 & 0.76\\

         Simple shot (conv4)  & Classes 1-13& 65.0$\pm$ 2.3\% & 0.65 & 0.65\\
         
         Simple shot (ResNet18) & ImageNet&  \textbf{85.2$\pm$2.0\%}&\textbf{0.86}&\textbf{0.85}\\
         kNN (k=1) on bare pixels & Classes 14-15&  71.7$\pm$2.6\%&0.72&0.72\\
         kNN (k=3) on bare pixels & Classes 14-15&  59.3$\pm$2.9\%&0.60&0.59\\
         \hline
        \end{tabular} 
    \caption{Classification accuracies, macro-averaged precisions, and macro-averaged recalls on different subsets of our full dataset (MLSTM) data for the 3-shot case. Accuracies averaged over 100 episodes and with 95\% confidence interval. Accuracies listed here are assuming that each class appears an equal number of times to give a fair representation of accuracy. Precision and recall are calculated cumulatively over the full 100 episodes. } 
    \label{tab: acc_abl_2_3_shot}
    \end{center}
\end{table*}

\begin{table*}[!htb]
\footnotesize
    \begin{center}
        \begin{tabular}{|| c | c | c | c | c || }         
        \hline
     \multicolumn{5}{ |c| }{Si(001):H:AsH$_3$ (test set 
 of classes 5-8) } \\
         \hline
         \multirow{2}{1em}{Model} & Training & Acc & \highlight{Prec} & \highlight{Recall} \\ 
          & \& validation data & (4-way, 1-shot) & (4-way, 1-shot)& (4-way, 1-shot)  \\
        \hline
         Prototypical &Classes $\pm$1-4, $\pm$9-15&  \textbf{ 93.6$\pm$ 0.9\%} & \textbf{0.94}&\textbf{0.94}\\ 
         
         Prototypical & Classes 1-4, 9-15 &  87.7$\pm$ 1.4\%& 0.88&0.88\\ 
         
         Matching  &Classes $\pm$1-4, $\pm$9-15&  84.3$\pm$1.4  \% & 0.87 & 0.84 \\
         
         Matching  &Classes 1-4, 9-15&  77.8$\pm$1.7  \% & 0.78 & 0.78 \\
         
         Relation &Classes $\pm$1-4, $\pm$9-15&  54.7 $\pm$ 2.0 \%& 0.56& 0.38\\
         
         Relation &Classes 1-4, 9-15&  50.65$\pm$1.8 \%& 0.57& 0.51\\
         
         Simple shot (conv4) &Classes $\pm$1-4, $\pm$9-15&  88.5$\pm$1.7\%& 0.89 & 0.89\\
         
         Simple shot (conv4) &Classes 1-4, 9-15&  80.4$\pm$1.6\%& 0.81 & 0.80 \\
         
         Simple shot (ResNet18) & ImageNet  &   77.8$\pm$ 1.6\%& 0.79&0.78 \\
         kNN (k=1) on bare pixels & Classes 5-8& 76.4$\pm$2.1 \% & 0.80 & 0.76\\
         \hline
     \multicolumn{5}{ |c| }{Ge(001):AsH$_3$ (test set of classes 9-13)} \\
         \hline
         \multirow{2}{1em}{Model} &Training &Acc & \highlight{Prec} & \highlight{Recall}\\
         & \& validation data & (4-way, 1-shot) & (4-way, 1-shot)& (4-way, 1-shot) \\
         \hline
         Prototypical & Classes $\pm$1-8, $\pm$14-15 & 54.5$\pm$2.0\%& \textbf{0.54}&\textbf{0.54}\\

         Prototypical & Classes 1-8, 14-15 & \textbf{54.8$\pm$2.1\%}& \textbf{0.54} & \textbf{0.54}\\
        
         Matching     & Classes $\pm$1-8, $\pm$14-15 & 50.1$\pm$ 2.1\%& 0.49 & 0.49\\

         Matching     & Classes 1-8, 14-15 & 42.8$\pm$ 2.1\% & 0.43 & 0.42\\
         
         Relation     & Classes $\pm$1-8, $\pm$14-15 &  38.9$\pm$1.7 \%& 0.40&0.39\\
         
         Relation     & Classes 1-8, 14-15 &  30.1$\pm$1.3 \%&  0.38&0.30\\
         
         Simple shot (conv4) & Classes $\pm$1-8, $\pm$14-15&  52.4$\pm$2.0 \%& 0.53& 0.52\\
         
         Simple shot (conv4) & Classes 1-8, 14-15&  51.8$\pm$2.0 \%&  0.52 & 0.51\\
         
         Simple shot (ResNet18) & ImageNet &   45.8$\pm$ 1.9\%& 0.46 &0.45\\
         kNN (k=1) on bare pixels & Classes 9-13& 49.1$\pm$ 1.5\%&  0.54&0.49\\
         kNN (k=3) on bare pixels & Classes 9-13& 45.9$\pm$2.1 \%& 48 & 46\\
         \hline
     \multicolumn{5}{ |c| }{TiO$_2$(110) (test set of classes 14-15)} \\
        \hline
         \multirow{2}{1em}{Model} &Training &Acc & \highlight{Prec} & \highlight{Recall}\\
         & \& validation data & (2-way, 1-shot) & (2-way, 1-shot) & (2-way, 1-shot) \\
        \hline
         Prototypical & Classes $\pm$1-13 & 66.4$\pm$ 2.3 \% & 0.69 & 0.63\\ 
         
         Prototypical & Classes 1-13 & 63.4$\pm$2.5 \% & 0.65 & 0.63\\ 
         
         Matching     & Classes $\pm$1-13&  63.8 $\pm$2.5\%&0.65&0.64\\
         
         Matching     & Classes 1-13&  61.9 $\pm$2.9\%& 0.63 & 0.62 \\
         
         Relation     & Classes $\pm$1-13 & 62.6$\pm$3.3 \% & 0.64 & 0.63 \\
         
         Relation     & Classes 1-13&  46.2$\pm$ 2.6\%& 0.46 & 0.46\\
         
         Simple shot (conv4)  & Classes $\pm$1-13&  69.2$\pm$3.1\%&   0.69&0.69 \\
         
         Simple shot (conv4)  & Classes 1-13&  61.3$\pm$2.5\%&   0.61&0.61 \\
         
         Simple shot (ResNet18) & ImageNet& \textbf{75.9$\pm$ 2.4\%}&  \textbf{0.80}&\textbf{0.76}\\
         kNN (k=1) on bare pixels & Classes 14-15&  57.5$\pm$2.9\%&  0.58&0.57\\
         \hline
        \end{tabular} 
    \caption{Classification accuracies, macro-averaged precisions, and macro-averaged recalls on different subsets of our full dataset (MLSTM) data for the 1-shot case. Accuracies averaged over 100 episodes and with 95\% confidence interval. Accuracies listed here are assuming that each class appears an equal number of times to give a fair representation of accuracy. Precision and recall are calculated cumulatively over the full 100 episodes.} 
    \label{tab: acc_abl_2_1_shot}
    \end{center}
\end{table*}

\clearpage

\subsection{Exploration into Accuracy Variation Due to Randomness Sources}\label{randomnedd_appendix}
\highlight{When training neural networks, some variability in performance between runs is expected. This arises from several sources of stochasticity, including random weight initialisation, the use of random number generators during data augmentation and data loading, and numerical effects associated with finite-precision (64-bit) arithmetic. In large-scale training regimes, such variability is typically limited to a few percentage points, as large datasets stabilise optimisation and promote convergence to similar minima. Since our study operates in a markedly low-data regime, we explicitly investigate how this affects the reproducibility of our results.}

\highlight{In the main manuscript, we report confidence intervals that capture variability arising from the random selection of support and query sets during evaluation. Here, we consider two additional sources of randomness: (i) random weight initialisation and (ii) the partitioning of data into training and validation sets. Variability due to random initialisation is summarised in Table~\ref{tab:acc_random_init}, while the effect of different data splits is reported in Table~\ref{tab:acc_k_fold}.}

\highlight{To assess the impact of random initialisation, each model was trained 50 times using different random seeds. Each trained model was then evaluated on the test set over 100 episodes. The resulting accuracies were averaged across runs, and the corresponding standard deviations were computed.}

\highlight{For the episodically trained models (prototypical, matching, and relation networks), variability in data splits was introduced by changing which entire classes were assigned to the training or validation sets. In contrast, the SimpleShot model is trained using a conventional supervised paradigm; accordingly, variability was introduced by altering which individual samples within each class were allocated to the training or validation sets. To quantify the effect of data partitioning, we generated 50 random splits of the dataset and trained a separate model for each split. Each model was evaluated in the same manner, and the mean accuracy and standard deviation were computed.}

\highlight{Among the trained models, the relation network consistently exhibits the largest variability, with standard deviations reaching up to 11\%. This increased sensitivity is likely attributable to its higher model complexity: unlike the other approaches, the relation network employs a learned distance function implemented as an additional neural network. The resulting increase in the number of trainable parameters makes the model more susceptible to stochastic effects during training, particularly in the low-data regime considered here.}

\highlight{Since we have trained many networks with different combinations of training and validation sets, and training each one of these 100 times to carry out this study, we have focused only on the ones that we present in the main body of the paper.}

\begin{table*}[!htb]
\footnotesize
    \begin{center}
        \begin{tabular}{|| c | c | c | c || }         
        \hline
     \multicolumn{4}{ |c| }{Si(001):H:AsH$_3$ (test set 
 of classes 5-8) } \\
         \hline
         \multirow{2}{1em}{Model} & Training & Acc &  Acc \\ 
          & \& validation data & (4-way, 1-shot) & (4-way, 3-shot)  \\
        \hline
         Prototypical &Classes $\pm$1-4, $\pm$9-15&  \textbf{95.2$\pm$0.8\%} &  \textbf{98.0$\pm$0.6\%} \\ 
         Matching  &Classes $\pm$1-4, $\pm$9-15&  74.3$\pm$4.1\% & 70.5$\pm$7.3\% \\
         Relation &Classes $\pm$1-4, $\pm$9-15&  80.6$\pm$8.0\%& 57.5$\pm$9.1\%\\
         Simple shot (conv4) &Classes $\pm$1-4, $\pm$9-15&  86.7$\pm$3.8\%& 93.5$\pm$2.4\% \\
         Simple shot (ResNet18) & ImageNet  &  N/A & N/A \\
         kNN (k=1) on bare pixels & Classes 5-8 & N/A & N/A \\      
         \hline
     \multicolumn{4}{ |c| }{Ge(001):AsH$_3$ (test set of classes 9-13)} \\
         \hline
         \multirow{2}{1em}{Model} &Training &Acc &  Acc \\
         & \& validation data & (4-way, 1-shot) &(4-way, 3-shot) \\
         \hline
         Prototypical & Classes $\pm$1-8, $\pm$14-15 & \textbf{53.2$\pm$1.0\%} & \textbf{59.8$\pm$0.9\%} \\ 
         Matching     & Classes $\pm$1-8, $\pm$14-15 & 42.2$\pm$2.0\% & 51.8$\pm$3.2\% \\
         Relation     & Classes $\pm$1-8, $\pm$14-15 & 39.6$\pm$2.9\% & 45.1$\pm$4.5\% \\
         Simple shot (conv4) & Classes $\pm$1-8, $\pm$14-15 & 49.6$\pm$2.2\% & 57.1$\pm$2.6\% \\
         Simple shot (ResNet18) & ImageNet & N/A & N/A\\
         kNN (k=1) on bare pixels & Classes 9-13& N/A & N/A \\
         \hline
     \multicolumn{4}{ |c| }{TiO$_2$(110) (test set of classes 14-15)} \\
        \hline
         \multirow{2}{1em}{Model} &Training &Acc &  Acc \\
         & \& validation data & (2-way, 1-shot) & (2-way, 3-shot) \\
        \hline
         Prototypical & Classes $\pm$1-13 & \textbf{67.5$\pm$0.4\%} & \textbf{67.6$\pm$1.6\%} \\ 
         Matching     & Classes $\pm$1-13 & 56.0$\pm$2.1\% & 62.5$\pm$3.1\% \\
         Relation     & Classes $\pm$1-13 & 63.0$\pm$7.8\% & 63.6$\pm$6.1\% \\
         Simple shot (conv4)  & Classes $\pm$1-13 & 61.5$\pm$2.2\% & 70.3$\pm$2.1\% \\
         Simple shot (ResNet18) & ImageNet& N/A & N/A \\
         kNN (k=1) on bare pixels & Classes 14-15& N/A & N/A \\
         \hline
        \end{tabular} 
\caption{\highlight{Classification accuracies for the different models, and their variation due to the random initialisation of the weights. Note that the ResNet18 and kNN models are not affected by this as they do not have a random initialisation of weights. The accuracies reported are averages over 50 random initialisations, with the standard deviation reported.}}
    \label{tab:acc_random_init}
    \end{center}
\end{table*}

\begin{table*}[!htb]
\footnotesize
    \begin{center}
        \begin{tabular}{|| c | c | c | c || }         
        \hline
     \multicolumn{4}{ |c| }{Si(001):H:AsH$_3$ (test set of classes 5-8) } \\
         \hline
         \multirow{2}{1em}{Model} & Training & Acc &  Acc \\ 
          & \& validation data & (4-way, 1-shot) & (4-way, 3-shot)  \\
        \hline
         Prototypical &Classes $\pm$1-4, $\pm$9-15&  \textbf{95.0$\pm$1.5\%} &  \textbf{97.6$\pm$1.0\%} \\ 
         Matching  &Classes $\pm$1-4, $\pm$9-15&  77.5$\pm$6.4\% & 79.2$\pm$6.8\% \\
         Relation &Classes $\pm$1-4, $\pm$9-15&  58.9$\pm$7.3\% & 54.4$\pm$11.1\%\\
         Simple shot (conv4) &Classes $\pm$1-4, $\pm$9-15&  87.7$\pm$3.0\% & 94.0$\pm$1.9\% \\
         Simple shot (ResNet18) & ImageNet  & N/A  & N/A\\
         kNN (k=1) on bare pixels & Classes 5-8&N/A & N/A \\      
         \hline
     \multicolumn{4}{ |c| }{Ge(001):AsH$_3$ (test set of classes 9-13)} \\
         \hline
         \multirow{2}{1em}{Model} &Training &Acc &  Acc \\
         & \& validation data & (4-way, 1-shot) &(4-way, 3-shot) \\
         \hline
         Prototypical & Classes $\pm$1-8, $\pm$14-15 & \textbf{52.6$\pm$1.6\%}& \textbf{63.6$\pm$2.7\%} \\
         Matching     & Classes $\pm$1-8, $\pm$14-15 & 46.0$\pm$4.1\% & 53.5$\pm$5.3\%\\
         Relation     & Classes $\pm$1-8, $\pm$14-15 & 32.8$\pm$5.4\% & 31.3$\pm$5.4\%\\
         Simple shot (conv4) & Classes $\pm$1-8, $\pm$14-15& 49.7$\pm$1.4\% & 57.3$\pm$1.8\%\\
         Simple shot (ResNet18) & ImageNet &   N/A& N/A \\
         kNN (k=1) on bare pixels & Classes 9-13& N/A & N/A \\
         \hline
     \multicolumn{4}{ |c| }{TiO$_2$(110) (test set of classes 14-15)} \\
        \hline
         \multirow{2}{1em}{Model} &Training &Acc &  Acc \\
         & \& validation data & (2-way, 1-shot) & (2-way, 3-shot) \\
        \hline
         Prototypical & Classes $\pm$1-13 & \textbf{67.9$\pm$1.0\%} & \textbf{69.0$\pm$2.4\%} \\ 
         Matching     & Classes $\pm$1-13&  57.9$\pm$4.0\%& 63.3$\pm$3.9\%\\
         Relation     & Classes $\pm$1-13&  55.3$\pm$7.0\%& 51.4$\pm$4.2\%\\
         Simple shot (conv4)  & Classes $\pm$1-13&  60.3$\pm$2.0\%& 69.3$\pm$1.4\%\\
         Simple shot (ResNet18) & ImageNet& N/A&N/A\\
         kNN (k=1) on bare pixels & Classes 14-15&  N/A& N/A\\
         \hline
        \end{tabular} 
\caption{\highlight{Classification accuracies for the different models, and their variation due to the different splits of the data. Note that the ResNet18 and kNN models are not affected by this as they do not have a training and validation set. The accuracies reported are averages over 50 different splits, with the standard deviation reported.}}
    \label{tab:acc_k_fold}
    \end{center}
\end{table*}

\clearpage

\subsection{Relation Network Modification} \label{relnet_appendix}
\begin{figure*}[!htbp]
    \centering
    \includegraphics[width=15cm]{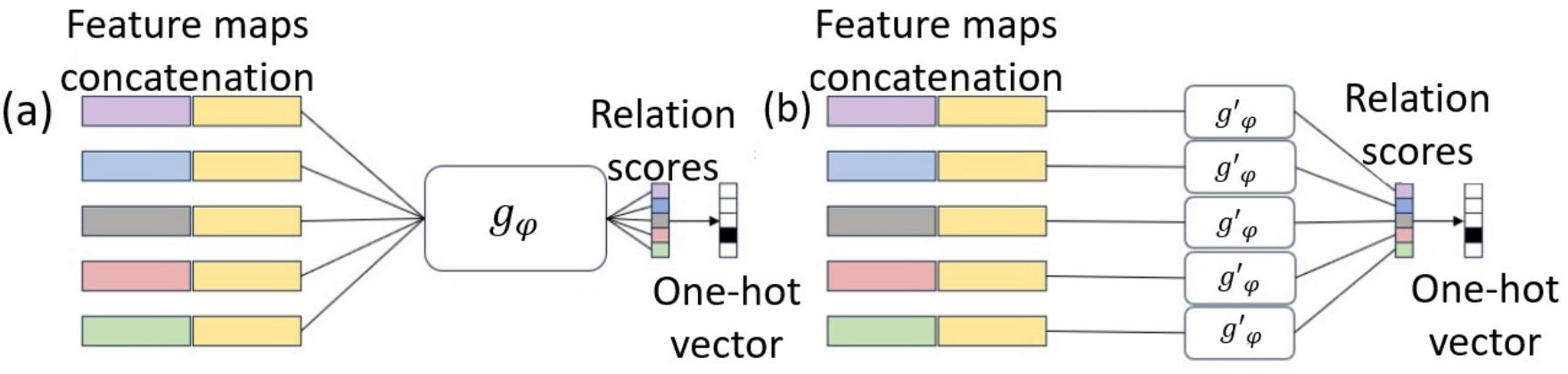}
    \caption{ (a) The relation module from the original relation network paper with a neural network metric $g_{\phi}$ \cite{relation_net}. The non-yellow blocks are the feature vectors of the query vectors. The support vector's feature vector is represented by the yellow blocks. (b) The relation module used in this work. $g'_{\phi}$ takes a single feature map concatenation at a time and outputs a relation score. This allows for a variable number of classes, N, and shots, K. The colour coding of the blocks corresponds to the same vectors as in (a). }
    \label{relation_module}
\end{figure*}

We introduce a slight modification to the relation network. In the original relation network paper \cite{relation_net}, the query feature vector is concatenated with each support vector and the full tensor of shape $(N, \ K, \ 2 \times d )$ is forwarded through the relation module at once, as shown in figure \ref{relation_module} (a) (where $d$ is the dimension of the feature space). This architecture does not allow us to change N. However, the number of species on a surface recorded by an STM will vary between experiments, meaning a variable N. The proposed modified architecture, finds the relation scores in a pairwise fashion for each query and support vector separately. These are then combined to give a prediction. To the best of our knowledge, there is no literature that shows a relation network that has previously been used in this way to allow for a variable N. Since the relation network's distance metric is a second neural network, called the relation module, it has more parameters than the other networks investigated. The original has $2NKd m_L \prod_{n=1}^{L-1} m_L + \sum_{n=1}^{L-1} m_i $ more, while the siamese relation network has $4Kd m_L \prod_{n=1}^{L-1} m_L + \sum_{n=1}^{L-1} m_i $ more (where $m_i$ are the number of nodes in each hidden layer of the relation module, and L is the number of hidden layers).

\clearpage

\subsection{Application of unsupervised segmentation to STEM data} \label{STEM_appendix}
\begin{figure*}[!htb]
    \centering
    \includegraphics[width=14cm, height=3cm]{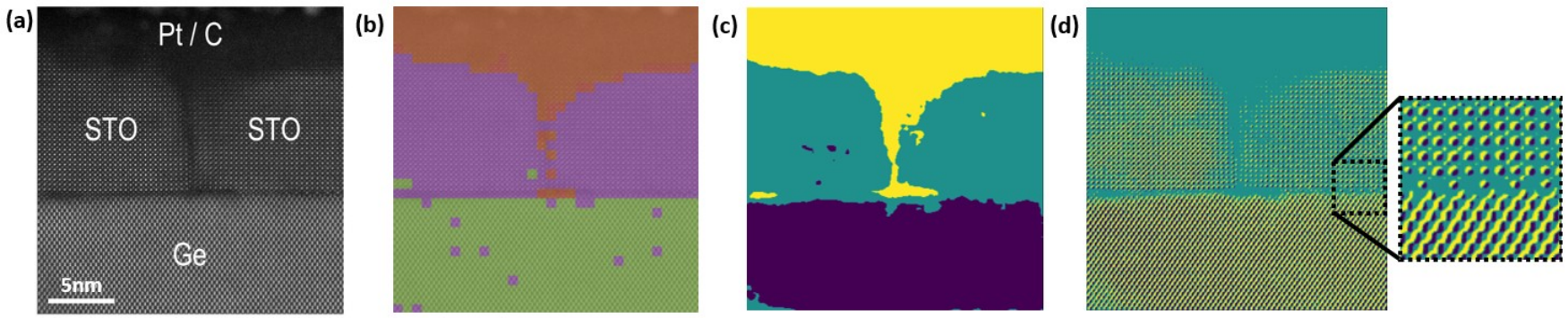}
    \caption{Comparison of the STEM data segmentation from \cite{akers_prototyp_SEM} with the output of our unsupervised segmentation technique applied to the same STEM data. (a) Original 3042$\times$3044 pixels STEM image of STO/Ge. (b) Segmentation of image from \cite{akers_prototyp_SEM}. (a) and (b) are reproduced from \cite{akers_prototyp_SEM} under a Creative Commons CC BY 4.0 license (https://creativecommons.org/licenses/by/4.0/). No changes were made. (c) Segmentation using technique from section 3.2 using a down sampled version that was 350$\times$350 pixels. (d) Segmentation using technique from section 3.2. Image here was down sampled to 1500$\times$1500 pixels. }
    
    \label{STEM}
\end{figure*}

The unsupervised labelling method described in section 3 may also be useful for other microscopy techniques. In Figure \ref{STEM} we display a comparison of the output from our method to that of reference \cite{akers_prototyp_SEM}, which describes course segmentation of STEM data using a prototypical network. Our technique can segment down to the pixel level, since we extract features for each pixel rather than a patch. Our approach allows users to vary the level of detail they desire, from phase separation to atomic level  segmentation. However, the technique in \cite{akers_prototyp_SEM} requires much less compute so is able to run on a CPU.

\twocolumn 

\end{document}